\documentclass[amsmath,amssymb,nofootinbib,notitlepage,superscriptaddress,twocolumn]{revtex4-1}
\pdfoutput=1

\usepackage{aas_macros,esint,overpic}
\usepackage[bottom]{footmisc}
\usepackage[colorlinks=true]{hyperref}
\usepackage[raggedright]{subfigure}
\let\originalleft\left
\let\originalright\right
\renewcommand{\left}{\mathopen{}\mathclose\bgroup\originalleft}
\renewcommand{\right}{\aftergroup\egroup\originalright}

\newcommand{\ab}[1]{\left|#1\right|}
\newcommand{\br}[1]{\left[#1\right]}
\newcommand{\cu}[1]{\left\{#1\right\}}
\newcommand{\pa}[1]{\left(#1\right)}

\newcommand{\dt}{\mathop{}\!\delta}
\newcommand{\ed}{\mathop{}\!\mathrm{d}}
\newcommand{\pd}{\mathop{}\!\partial}

\DeclareMathOperator\arctanh{arctanh}
\DeclareMathOperator\cd{cd}
\DeclareMathOperator\sn{sn}
\DeclareMathOperator\sign{sign}

\begin{document}

\title{Lensing by Kerr Black Holes}

\author{Samuel E. Gralla}
\email{sgralla@email.arizona.edu}
\affiliation{Department of Physics, University of Arizona, Tucson, Arizona 85721, USA}
\author{Alexandru Lupsasca}
\email{lupsasca@fas.harvard.edu}
\affiliation{Center for the Fundamental Laws of Nature, Harvard University, Cambridge, Massachusetts 02138, USA}
\affiliation{Society of Fellows, Harvard University, Cambridge, Massachusetts 02138, USA}

\begin{abstract}
Interpreting horizon-scale observations of astrophysical black holes demands a general understanding of null geodesics in the Kerr spacetime.  These may be divided into two classes: ``direct'' rays that primarily determine the observational appearance of a given source, and highly bent rays that produce a nested sequence of exponentially demagnified images of the main emission---the so-called ``photon ring''.  We develop heuristics that characterize the direct rays and study the highly bent geodesics analytically.  We define three critical parameters $\gamma$, $\delta$, and $\tau$ that respectively control the demagnification, rotation, and time delay of successive images of the source, thereby providing an analytic theory of the photon ring.  These observable parameters encode universal effects of general relativity, independent of the details of the emitting matter.
\end{abstract}

\maketitle

\section{Introduction}

With the advent of horizon-scale observations of astrophysical black holes \cite{EHT2019a,EHT2019b,EHT2019c,EHT2019d,EHT2019e,EHT2019f}, the intricate properties of null geodesics in the Kerr spacetime \cite{Carter1968,Bardeen1973,Vazquez2004,James2015,KerrGeodesics} are fast becoming a matter of practical relevance to astronomy.  Thanks to ray-tracing codes now operating with exquisite accuracy and speed \cite{Chan2013,Dexter2016,Moscibrodzka2018}, determining the observational appearance of a specified emission model is a quick and routine task.  However, given the enormous uncertainty in the nature of the emission arising from the present targets M87* and Sgr~A*, the ``inverse problem'' may be more relevant: Given an observation, what can one learn about the emission profile of the source?

Answering this kind of question demands a \textit{general} understanding of the effects of gravitational lensing in the Kerr spacetime.  The authors of the present manuscript have been involved in separate, recent efforts in this direction \cite{Gralla2019,Johnson2019}.  Reference~\cite{Gralla2019} argued that bright rings of emission from optically thin matter \cite{Jaroszynski1997,Falcke2000,Johannsen2010}  (hereafter, ``photon rings''\footnote{We use the term ``photon ring'' to describe the collection of demagnified images that appear near a closed curve on the image plane.  When optically thin matter emits from the vicinity of the black hole, these images superpose to provide a brightness enhancement.   We use the name ``critical curve'' for the curve where the images accumulate, and avoid the word ``shadow'' altogether.}) should be understood as superposed, exponentially demagnified images of the main emission, predicting a distinctive multipeak structure and giving the first quantitative estimate of the typical brightness enhancement (a factor of about 2--3).  Soon after, Ref.~\cite{Johnson2019} obtained a formula for the asymptotic demagnification factor as a function of black hole spin and observer inclination, confirmed the typical brightness enhancement and multipeak structure in state-of-the-art models \cite{EHT2019e} ray-traced at higher resolution than previously considered, and proposed an experimental method for detecting the discrete peaks using space-based interferometry.  In this paper, we unite our perspectives on the problem and significantly generalize these results, with the aim of presenting a complete guide to understanding lensing by Kerr black holes.

We have developed two new analytic tools in service of this goal: 1) a complete, fully explicit solution of the Kerr null geodesic equation expressed in terms of Legendre elliptic integrals and Jacobi elliptic functions (presented in a companion paper \cite{KerrGeodesics}), and 2) a logarithmic approximation valid for highly bent photons (derived in App.~\ref{app:MAE}).   We use the first tool to explore general properties of null geodesics, and exploit the second to provide a detailed analytic theory of the photon ring.

It is helpful to organize the analysis by the number of orbits that an emitted photon executes before reaching the detector (Fig.~\ref{fig:Backside}).  For ``direct'' photons that complete of order half-an-orbit or less, we find that the spin of the black hole has little influence on the trajectory.  For example, we show that for an equatorial (i.e., spin-aligned or antialigned) disk of emission viewed face-on, the arrival impact parameter $b$ of a photon emitted from Boyer-Lindquist radius $r_s$ is given by ``just adding one'',
\begin{align}
	\label{eq:DirectHeuristics}
	\frac{b}{M}\approx\frac{r_s}{M}+1,
\end{align}
with this formula holding empirically to $10\%$ accuracy at all spin (Fig.~\ref{fig:Transfer} left).  For observers inclined relative to the disk, the spin still has little effect on the arrival position from a fixed equatorial radius, although it does shrink the apparent size of the equator of the black hole (Fig.~\ref{fig:Contours}).  For models with emission extending to the horizon, the observed central dark area will correspondingly shrink.

Photons that make of order half-an-orbit to one orbit contribute a demagnified image of the source over a band surrounding a critical curve on the image plane.  For a diffuse, optically thin source near the horizon, this image superposes onto the direct emission to produce a thin ring with diameter ${\sim}10M$, width ${\sim}M$, and about twice the background intensity \cite{Gralla2019},\footnote{This image was called the ``lensing ring'' in Ref.~\cite{Gralla2019}; here, we follow Ref.~\cite{Johnson2019} and include it as part of the ``photon ring''.} a striking feature in simulated images \cite{EHT2019e}.  Here, we show that the precise width of this band varies significantly with spin (Fig.~\ref{fig:Transfer} left), especially in the region corresponding to photons emitted from the vicinity of the horizon.  For models with emission near the horizon, the demagnified image will therefore be broader, and contribute significantly more flux, when the black hole spins rapidly.

Photons executing of order one orbit or more contribute a sequence of highly demagnified images near the critical curve \cite{Luminet1979,Beckwith2005,Gralla2019,Johnson2019}.  We derive an asymptotic expansion for the number of orbits as a function of the (perpendicular) distance from the critical curve, and show that the resulting logarithmic approximation is excellent even for photons executing of order only a single orbit (Fig.~\ref{fig:Screen}).  We develop a precise analytic theory of the demagnified images based on three key quantities defined for the bound photon orbits:
\begin{itemize}
	\item The Lyapunov exponent $\gamma$ characterizing the instability of the bound orbit, defined relative to a half-libration in polar angle $\theta$ \cite{Johnson2019}.
	\item The change $\delta$ in azimuthal angle $\phi$ over a polar half-libration \cite{Teo2003}.
	\item The period $\tau$ of a polar half-libration.
\end{itemize}
We show that for an equatorial disk of emission viewed face-on, each successive image is demagnified by a factor of $e^{-\gamma}$, rotated by an angle $\delta$, and delayed by a time $\tau$.  These images alternate between showing the front side and the backside of the disk (Fig.~\ref{fig:Backside}).  For nonequatorial sources, we instead distinguish two families of images, each with demagnification $e^{-2\gamma}$, rotation $2\delta$, and time delay $2\tau$.  These simple associations break down when the observer is significantly inclined, but we are still able to make precise statements about the origin of emission as a function of observed position near the critical curve.  

These results unite and generalize our previous treatments of the demagnification factor \cite{Gralla2019,Johnson2019}, while also introducing $\delta$ and $\tau$ as additional key quantities characterizing the demagnified images.  The spin-dependent critical parameters $\gamma$, $\delta$ and $\tau$ control universal (matter-independent) features of general relativity that could in principle be observed with future detectors.

This paper is organized as follows.  In Sec.~\ref{sec:Framework}, we review and present a useful formalism for Kerr null geodesics.  Next, in Sec.~\ref{sec:CriticalRays}, we analyze the bound photon orbits, and define their critical parameters $\gamma$, $\delta$, and $\tau$.  Then, in Sec.~\ref{sec:Observer}, we discuss the screen of a distant observer, presenting new details about the map from conserved quantities to position in the image plane.  We describe properties of complete rays in the Kerr exterior in Sec.~\ref{sec:Rays}, and study segments of rays that represent propagation from source to observer in Sec.~\ref{sec:Photons}.  Finally, in Sec.~\ref{sec:PhotonRing}, we develop the analytic theory of the photon ring in terms of the critical parameters $\gamma$, $\delta$, and $\tau$.

\section{General framework}
\label{sec:Framework}

We work with Boyer-Lindquist coordinates $(t,r,\theta,\phi)$ on the spacetime of a Kerr black hole with mass $M$ and angular momentum $J=Ma$, and define
\begin{align}
	\Sigma(r,\theta)=r^2+a^2\cos^2{\theta},\quad
	\Delta(r)=r^2-2Mr+a^2.
\end{align}
The roots of $\Delta(r)$ correspond to the outer/inner horizons
\begin{align}
	r_\pm=M\pm\sqrt{M^2-a^2}.
\end{align}
We assume that $0<a<M$, such that the coordinate $\phi$ increases in the sense of rotation of the black hole.  The nonrotating ($a\to0$) and extremal ($a\to M$) limits  may be taken after final observables are computed.

In discussing null geodesics, we will make a distinction between ``rays'' and ``photons''.  By a \textit{ray}, we will mean a complete null geodesic in the Kerr exterior, which enters from the white hole or the celestial sphere, before eventually leaving via the black hole or the celestial sphere.  By a \textit{photon}, we will mean a portion of a ray, which represents the emission and absorption (or observation) of light.  In radiative transport, one considers rays that propagate through a medium, gaining and losing photons (according to the local emissivity and absorptivity) on their way to the detector.

We will adopt the ``integral'' approach to the study of null geodesics in the Kerr spacetime.  In this approach, pioneered by Carter \cite{Carter1968} and Bardeen \cite{Bardeen1973}, one reduces the equations to quadratures using conserved quantities.  Building on important earlier developments \cite{Chandrasekhar1983,Rauch1994,Dexter2009,Kapec2019}, in a companion paper \cite{KerrGeodesics} we have classified all motions, reduced all integrals to real elliptic form, and inverted the equations to provide explicit, parameterized trajectories.  Herein, we only summarize the results needed for this paper; complete derivations may be found in Ref.~\cite{KerrGeodesics}.

Each Kerr photon trajectory possesses two conserved quantities $\lambda$ and $\eta$, corresponding to the energy-rescaled angular momentum and Carter integral, respectively.  These allow the four-momentum $p^\mu$ along the trajectory to be reconstructed as
\begin{subequations}
\label{eq:GeodesicEquation}
\begin{align}
	\frac{\Sigma}{E}p^r&=\pm_r\sqrt{\mathcal{R}(r)},\\
	\frac{\Sigma}{E}p^\theta&=\pm_\theta\sqrt{\Theta(\theta)},\\
	\frac{\Sigma}{E}p^\phi&=\frac{a}{\Delta}\pa{r^2+a^2-a\lambda}+\frac{\lambda}{\sin^2{\theta}}-a,\\
	\frac{\Sigma}{E}p^t&=\frac{r^2+a^2}{\Delta}\pa{r^2+a^2-a\lambda}+a\pa{\lambda-a\sin^2{\theta}},
\end{align}
\end{subequations}
where $E=-p_t$ is the constant ``energy at infinity''\footnote{We exclude the measure-zero set of geodesics with $E=0$ exactly.  In particular, such geodesics cannot reach an observer at infinity.} and 
\begin{align}
	\label{eq:RadialPotential}
	\mathcal{R}(r)&=\pa{r^2+a^2-a\lambda}^2-\Delta(r)\br{\eta+\pa{\lambda-a}^2},\\
	\Theta(\theta)&=\eta+a^2\cos^2{\theta}-\lambda^2\cot^2{\theta}.
\end{align}
The symbols $\pm_r$ and $\pm_\theta$ indicate the sign of $p^r$ and $p^\theta$, respectively.  Turning points in $r$ and $\theta$ occur at zeros of the radial and angular ``potentials'' $\mathcal{R}(r)$ and $\Theta(\theta)$, respectively.

Consider a null geodesic connecting spacetime events $(t_s,r_s,\theta_s,\phi_s)$ and $(t_o,r_o,\theta_o,\phi_o)$, where $s$ and $o$ stand for source and observer.  By integrating along the trajectory, the geodesic equation \eqref{eq:GeodesicEquation} may be recast in integral form,\footnote{We identify $\phi\sim\phi+2\pi$, allowing $\Delta\phi=\phi_o-\phi_s$ to take any value.  If we had instead restricted $\phi$ to lie within the canonical range $[0,2\pi)$, then the right-hand side of Eq.~\eqref{eq:IntegralGeodesicEquation1} would have to contain mod~$2\pi$.}
\begin{subequations}
\label{eq:IntegralGeodesicEquation}
\begin{align}
	\label{eq:IG}
	I_r&=G_\theta,\\
	\label{eq:IntegralGeodesicEquation1}
	\Delta\phi:=\phi_o-\phi_s&=I_\phi+\lambda G_\phi,\\
	\label{eq:IntegralGeodesicEquation2}
	\Delta t:=t_o-t_s&=I_t+a^2G_t,
\end{align}
\end{subequations}
where we define
\begin{subequations}
\label{eq:GeodesicIntegrals}
\begin{align}
	I_r&=\fint_{r_s}^{r_o}\frac{\ed r}{\pm_r\sqrt{\mathcal{R}(r)}},\\
	G_\theta&=\fint_{\theta_s}^{\theta_o}\frac{\ed\theta}{\pm_\theta\sqrt{\Theta(\theta)}},\\
	I_\phi&=\fint_{r_s}^{r_o}\frac{a\pa{2Mr-a\lambda}}{\pm_r\Delta(r)\sqrt{\mathcal{R}(r)}}\ed r,\\
	G_\phi&=\fint_{\theta_s}^{\theta_o}\frac{\csc^2{\theta}}{\pm_\theta\sqrt{\Theta(\theta)}}\ed\theta,\\
	I_t&=\fint_{r_s}^{r_o}\frac{r^2\Delta(r)+2Mr\pa{r^2+a^2-a\lambda}}{\pm_r\Delta(r)\sqrt{\mathcal{R}(r)}}\ed r,\\
	G_t&=\fint_{\theta_s}^{\theta_o}\frac{\cos^2{\theta}}{\pm_\theta\sqrt{\Theta(\theta)}}\ed\theta.
\end{align}
\end{subequations}
Here, the notation $\fint$ indicates that these integrals are to be understood as path integrals along the photon trajectory, with the signs $\pm_r=\sign(p^r)$ and $\pm_\theta=\sign(p^\theta)$ switching at radial and angular turning points, respectively.  In particular, all path integrals increase monotonically along the trajectory.

\subsection{Angular integrals}

The analysis of the angular integrals differs depending on the region of conserved quantity space.  In this paper, unless otherwise specified, we will restrict to positive $\eta$,
\begin{align}
	\eta>0,
\end{align}
thereby excluding the so-called ``vortical'' geodesics with $\eta<0$.  (This excludes only a small portion near the middle of an observer's screen, where the image is normally dark---see Fig.~\ref{fig:2-1} below.  Furthermore, equatorial sources cannot emit vortical photons, as these never intersect the equatorial plane.)  The $\eta>0$ geodesics librate between turning points $\theta_\pm$ above and below the equatorial plane,
\begin{align}
	\label{eq:TurningPoints}
	\theta_\pm=\arccos\pa{\mp\sqrt{u_+}},
\end{align}
where
\begin{align}
	u_\pm=\triangle_\theta\pm\sqrt{\triangle_\theta^2+\frac{\eta}{a^2}},\quad
	\triangle_\theta=\frac{1}{2}\pa{1-\frac{\eta+\lambda^2}{a^2}}.
\end{align}
To aid in the expression of the angular path integrals $G_\theta$, $G_\phi$, and $G_t$, we introduce the notation
\begin{align}
	F_i&=F\pa{\arcsin\pa{\frac{\cos{\theta_i}}{\sqrt{u_+}}}\left|\frac{u_+}{u_-}\right.},\\
	\Pi_i&=\Pi\pa{u_+;\arcsin\pa{\frac{\cos{\theta_i}}{\sqrt{u_+}}}\left|\frac{u_+}{u_-}\right.},\\
	E_i'&=E'\pa{\arcsin\pa{\frac{\cos{\theta_i}}{\sqrt{u_+}}}\left|\frac{u_+}{u_-}\right.},
\end{align}
where $i\in\cu{s,o}$ can be either source or observer.  Here, $F(\varphi|k)$, $E(\varphi|k)$, and $\Pi(n;\varphi|k)$ respectively denote the incomplete elliptic integrals of the first, second, and third kind,\footnote{Our conventions for elliptic integrals are listed in App.~A of Ref.~\cite{Kapec2019} and match the built-in implementation in \textit{Mathematica 12}.} while the prime denotes a derivative with respect to $k$, $E'(\varphi|k):=\pd_kE(\varphi|k)=\br{E(\varphi|k)-F(\varphi|k)}/(2k)$.  These integrals vanish at the equator,
\begin{align}
	F_i=\Pi_i=E_i'=0,\qquad
	\pa{\theta_i=0}
\end{align}
and become complete at turning points,
\begin{align}
	F_i=\mp K,\quad
	\Pi_i=\mp\Pi,\quad
	E_i'=\mp E,\qquad
	\pa{\theta_i=\theta_\pm}
\end{align}
where our notation for the complete elliptic integrals is
\begin{align}
	K&=K\pa{\frac{u_+}{u_-}}
	=F\pa{\frac{\pi}{2}\left|\frac{u_+}{u_-}\right.},\\ 
	\Pi&=\Pi\pa{u_+\left|\frac{u_+}{u_-}\right.}
	=\Pi\pa{u_+;\frac{\pi}{2}\left|\frac{u_+}{u_-}\right.},\\
	E'&=E'\pa{\frac{u_+}{u_-}}
	=E'\pa{\frac{\pi}{2}\left|\frac{u_+}{u_-}\right.}.
\end{align}
The $\eta>0$ angular path integrals may be written in terms of these quantities and the number $m$ of angular turning points encountered along the trajectory as \cite{Kapec2019,KerrGeodesics}
\begin{align}
	\label{eq:GthetaIntegral}
	G_\theta&=\frac{1}{a\sqrt{-u_-}}\br{2mK\pm_sF_s\mp_oF_o}, \\
	\label{eq:GphiIntegral}
	G_\phi&=\frac{1}{a\sqrt{-u_-}}\br{2m\Pi\pm_s\Pi_s\mp_o\Pi_o},\\
	\label{eq:GtIntegral}
	G_t&=-\frac{2u_+}{a\sqrt{-u_-}}\br{2mE'\pm_sE_s'\mp_oE_o'},
\end{align}
with $\pm_i$ denoting the sign of $p^\theta$ at the source ($i=s$) or observer ($i=o$) point,
\begin{align}
	\pm_i=\sign\pa{p_i^\theta}.
\end{align}
Since $p^\theta$ changes sign after each turning point, these signs obey the constraint
\begin{align}
	\pm_s=\pm_o(-1)^m.
\end{align}

Finally, note that the integral for $G_\theta$ can be inverted to solve for $\theta_o$ or $\theta_s$ as a function of $G_\theta$ \cite{Dexter2009,Kapec2019,KerrGeodesics}.  Since in this paper, we mainly fix the observer point (a telescope at infinity), we present $\theta_s$ in terms of $\theta_o$ and $G_\theta$.  This may be inferred from expressions for $\theta_o(G_\theta,\theta_s)$ by interchanging $s$ and $o$, before sending $G_\theta\to-G_\theta$ to compensate.\footnote{The future-directed geodesic from source to observer is also a past-directed geodesic from observer to source.  The path integrals are monotonically \textit{decreasing} for the past-directed geodesic, so after interchanging $s\leftrightarrow o$, we must also send $G_\theta\to-G_\theta$.}  From Eq.~(71) of Ref.~\cite{KerrGeodesics} (noting that $\tau$ therein denotes $G_\theta$, while $\nu_\theta$ therein denotes $\pm_s$), we find
\begin{align}
	\label{eq:SourceAngle}
	\frac{\cos\theta_s}{\sqrt{u_+}}=\sn\pa{F_o\pm_o\sign\pa{\eta}a\sqrt{-u_-}G_\theta\left|\frac{u_+}{u_-}\right.},
\end{align}
where $\sn(\varphi|k)$ denotes the Jacobi elliptic sine function.  This formula holds regardless of the sign of $\eta$ \cite{Kapec2019,KerrGeodesics}.

\subsection{Radial integrals}

In this paper, we will consider a distant observer at
\begin{align}
	r_o\to\infty.
\end{align}
Geodesics that reach this far observer have at most one radial turning point outside the horizon.  Given a choice of conserved quantities $(\lambda,\eta)$, a simple way to test whether the ray has a turning point is to compute $r_4(\lambda,\eta)$ via Eq.~\eqref{eq:r4} below.  If $r_4$ is real and outside the horizon, then the ray has a turning point at radius $r_4$; otherwise, the ray never encounters a turning point.

For the rays with no turning point, the radial integrals $I_r$, $I_\phi$, and $I_t$ are single-valued functions of $r_s$, while for the rays with a turning point, these radial integrals must be double-valued in order to track whether or not the turning point has been reached.  We will denote the number of turning points of a \textit{photon} (portion of a ray) by $w\in\cu{0,1}$.  The radial integral $I_r$ may then be written
\begin{align}
 	\label{eq:IrIntegral}
	I_r=\int_{r_s}^{\infty}\frac{\ed r}{\sqrt{\mathcal{R}(r)}}+2w\int_{r_4}^{r_s}\frac{\ed r}{\sqrt{\mathcal{R}(r)}},
\end{align}
and likewise for $I_\phi$ and $I_t$ with the appropriate integrands.\footnote{The integral $I_t$ will diverge as $r_o\to\infty$, so one should let $r_o\to\infty$ only after an observable is computed.}  We may relate $w$ to the emission direction by
\begin{align}
	\label{eq:w}
	w=
	\begin{cases}
		0&p^r_s>0,\\
		1&p^r_s<0\text{ (and }r_+<r_4<r_s).
	\end{cases}
\end{align}
The conditions $r_s>r_4>r_+$ ensure that $r_s$ lies along a ray that reaches infinity after passing through a turning point $r_4$ (such that negative initial radial momentum is allowed).  If these conditions are not both satisfied for a given choice of conserved quantities, then only $w=0$ is allowed for those quantities, i.e., only photons emitted outward will reach infinity.

A ray reaching infinity originates either from the event horizon (of the white hole) or from infinity.  We denote the associated radial integral $I_r$ by $I_r^\mathrm{total}$,
\begin{align}
	\label{eq:IrTotal}
	I_r^\mathrm{total}=
	\begin{cases} 
		\displaystyle2\int_{r_4}^{\infty}\frac{\ed r}{\sqrt{\mathcal{R}(r)}}&r_+<r_4\in\mathbb{R},\vspace{2pt}\\
		\displaystyle\int_{r_+}^{\infty}\frac{\ed r}{\sqrt{\mathcal{R}(r)}}&\text{otherwise},
       \end{cases}
\end{align}
where we remind the reader that a ray reaching infinity began at infinity if $r_4(\lambda,\eta)$ is real and greater than the horizon, and otherwise began at the (white hole) horizon.

The full set of radial integrals were evaluated and reduced to elliptic form in Ref.~\cite{KerrGeodesics}, building on previous work in Refs.~\cite{Rauch1994,Dexter2009}.  The necessary antiderivatives for computing Eqs.~\eqref{eq:IrIntegral} and \eqref{eq:IrTotal} are given in App.~\ref{app:RadialAnalysis} below.

As in Eq.~\eqref{eq:SourceAngle} for $\theta_s(G_\theta)$, one may derive an inversion formula for $r_s(I_r)$  \cite{Dexter2009,KerrGeodesics}.  Eq.~(B119) of Ref.~\cite{KerrGeodesics} gives a formula for $r_o$, and we may infer the formula for $r_s$ as described above Eq.~\eqref{eq:SourceAngle}, i.e., by interchanging $o$ and $s$ and then sending $I_r\to-I_r$.  Noting that $\tau=I_r$ therein, and letting $r_o\to\infty$, the emission radius is given by 
\begin{align}
	\label{eq:SourceRadius}
	r_s=\frac{r_4r_{31}-r_3r_{41}\sn^2\pa{\frac{1}{2}\sqrt{r_{31}r_{42}}I_r-\mathcal{F}_o\big|k}}{r_{31}-r_{41}\sn^2\pa{\frac{1}{2}\sqrt{r_{31}r_{42}}I_r-\mathcal{F}_o\big|k}},
\end{align}
with
\begin{align}
	\mathcal{F}_o=F\pa{\left.\arcsin{\sqrt{\frac{r_{31}}{r_{41}}}}\right|k},\quad
	k=\frac{r_{32}r_{41}}{r_{31}r_{42}}.
\end{align}
Here, we introduced the notation
\begin{align}
	r_{ij}=r_i-r_j,
\end{align}
with the roots $\cu{r_1,r_2,r_3,r_4}$ given in Eqs.~\eqref{eq:RadialRoots} below.  This formula is contingent on the radial integral $I_r$ being in the allowed range,
\begin{align}
	\label{eq:IrRange}
	0<I_r<I_r^\mathrm{total}.
\end{align}
Provided that Eq.~\eqref{eq:IrRange} is satisfied, Eq.~\eqref{eq:SourceRadius} gives the emission radius of a photon reaching infinity with conserved quantities $(\lambda,\eta)$.  This formula holds even when (some of) the radial roots are complex \cite{KerrGeodesics}.

\subsection{Fractional number of orbits}
\label{sec:Orbits}

It is useful to have some measure of the total number of orbits executed by a given photon.  However, since the spatial trajectory is three-dimensional, there is some arbitrariness in the definition of an orbit.  As in Ref.~\cite{Johnson2019}, we define the journey from the equator to a polar turning point $\theta_\pm$ to be one quarter of an orbit, so that beginning and ending at the same turning point constitutes one full orbit.  For a measure of the fractional number of orbits, we seek a quantity that grows monotonically from zero, increasing by 1 after completing an orbit as defined above.  Since the path integral $G_\theta$ satisfies the requisite monotonicity property, we simply normalize by its value $G_\theta^1$ over one orbit,\footnote{A quarter orbit contributes $G_\theta^{1/4}\sim K(u_+/u_-)$, in accordance with the name ``quarter period'' given to the elliptic integral $K(x)$ in the study of pendulum motion, which is precisely of the form \eqref{eq:SourceAngle}.} defining the fractional number of orbits $n$ to be
\begin{align}
	n=\frac{G_\theta}{G^1_\theta},
\end{align}
with
\begin{align}
	\label{eq:G1}
	G^1_\theta=2\int_{\theta_-}^{\theta_+}\frac{\ed\theta}{\sqrt{\Theta(\theta)}}
	=\frac{4K}{a\sqrt{-u_-}}.
\end{align}
Using $I_r=G_\theta$ [Eq.~\eqref{eq:IG} above], we equivalently have
\begin{align}
	\label{eq:n}
	n=\frac{a\sqrt{-u_-}}{4K}I_r.
\end{align}
Note that $I_r=G_\theta$ is also the Mino time parameter \cite{Mino2003} that decouples the differential equations \eqref{eq:GeodesicEquation}.  Our parameter $n$ is proportional to the Mino time and provides a new physical interpretation of this quantity.

\section{Critical rays}
\label{sec:CriticalRays}

For generic values of $\lambda$ and $\eta$, the radial potential \eqref{eq:RadialPotential} possesses four distinct roots \eqref{eq:RadialRoots}, of which the real subset corresponds to radial turning points.  At special ``critical'' values $\tilde{\lambda}$ and $\tilde{\eta}$, the radial potential may develop a double root at some special radius $\tilde{r}$,
\begin{align}
	\label{eq:DoubleRoot}
	\mathcal{R}(\tilde{r})=\mathcal{R}'(\tilde{r})=0.
\end{align}
This occurs for $\tilde{r}>r_+$ if \cite{Bardeen1973} and only if \cite{KerrGeodesics} 
\begin{align}
	\label{eq:lambdaCritical}
	\tilde{\lambda}&=a+\frac{\tilde{r}}{a}\br{\tilde{r}-\frac{2\tilde{\Delta}}{\tilde{r}-M}},\\
	\label{eq:etaCritical}
	\tilde{\eta}&=\frac{\tilde{r}^{3}}{a^2}\br{\frac{4M\tilde{\Delta}}{\pa{\tilde{r}-M}^2}-\tilde{r}},
\end{align}
where $\tilde{r}$ must lie in the range $\tilde{r}\in\br{\tilde{r}_-,\tilde{r}_+}$, with
\begin{align}
	\label{eq:PhotonShell}
	\tilde{r}_\pm=2M\br{1+\cos\pa{\frac{2}{3}\arccos\pa{\pm\frac{a}{M}}}}.
\end{align}
Here and below, we use the notation $\tilde{Q}$ for a quantity $Q$ evaluated at criticality, i.e., at $r=\tilde{r}$, $\lambda=\tilde{\lambda}$, and $\eta=\tilde{\eta}$.

The double root \eqref{eq:DoubleRoot} indicates the existence of orbits with fixed Boyer-Lindquist radius $\tilde{r}$, i.e., \textit{bound photon orbits}.  At the boundaries \eqref{eq:PhotonShell} of the allowed range, the orbits are circular, equatorial, and prograde ($\tilde{r}_-$) or retrograde ($\tilde{r}_+$), whereas for intermediate radii the orbits also librate between turning points $\theta_-$ and $\theta_+$ given in Eq.~\eqref{eq:TurningPoints} above [note from Eqs.~\eqref{eq:etaCritical} and \eqref{eq:PhotonShell} that $\tilde{\eta}\geq0$].  The pole-crossing orbits $\tilde{\lambda}=0$ (where the turning points approach the poles) lie at the radius $\tilde{r}=\tilde{r}_0$ given by
\begin{align}
	\label{eq:CriticalRadius}
	\tilde{r}_0&=M+2\sqrt{M^2-\frac{a^2}{3}}\cos\br{\frac{1}{3}\arccos{\frac{\pa{1-\frac{a^2}{M^2}}}{\pa{1-\frac{a^2}{3M^2}}^{3/2}}}}.
\end{align}

Thus, the region of the Kerr spacetime spanned by bound photon orbits takes the shape of a spherical shell of variable thickness (the ``photon shell''), which is thickest at the equator and vanishingly thin at the pole (e.g., Fig.~2 of Ref.~\cite{Johnson2019}).  This shell is largest in the extremal limit $a\to M$, in which its range extends from $\tilde{r}_-=M$ to $\tilde{r}_+=4M$ at the equator.  In the nonrotating limit $a\to0$, the shell is vanishingly thin everywhere, degenerating to the ``photon sphere'' $r=3M$.  

Since there are no orbits that oscillate between two radial turning points outside the horizon, the bound photon orbits are unstable.  The rate of deviation of nearby orbits may be characterized by a Lyapunov exponent, which is usually defined with respect to a coordinate or affine time (e.g., as in Ref.~\cite{Yang2012}).  We will instead follow  Ref.~\cite{Johnson2019} and define the exponent using the fractional number of orbits as a parameter.  Consider a precisely critical ray with conserved quantities $\tilde{\lambda}(\tilde{r})$ and $\tilde{\eta}(\tilde{r})$, but that is not precisely at the radius $\tilde{r}$.  (Such rays approach the critical radius in the asymptotic future or past.)  In the regime $\ab{r-\tilde{r}}\ll\tilde{r}$, a simple calculation (App.~A1 of \cite{Johnson2019}) gives\footnote{By including a factor of 2, we are effectively defining the Lyapunov exponent with respect to the fractional number of \textit{half-}orbits, $2n$.  This choice was made for consistency with Ref.~\cite{Johnson2019}; note, however, that Ref.~\cite{Johnson2019} used the letter $n$ to denote the fractional number of half-orbits, whereas we have instead followed Ref.~\cite{Gralla2019} in using $n$ for the fractional number of orbits.}
\begin{align}
	\frac{r_2-\tilde{r}}{r_1-\tilde{r}}\approx e^{2\gamma\pa{n_2-n_1}},
\end{align}
where $r_1$ and $r_2$ denote the photon radius after executing $n_1$ and $n_2$ fractional orbits, respectively, while the Lyapunov exponent is 
\begin{align}
	\label{eq:LyapunovExponent}
	\gamma=\frac{4\tilde{r}\sqrt{\tilde{\chi}}}{a\sqrt{-\tilde{u}_-}}\tilde{K}.
\end{align}
Here, $\tilde{K}=K(\tilde{u}_+/\tilde{u}_-)$ is evaluated using the critical conserved quantities according to the convention established above, while $\tilde{\chi}$ is defined as
\begin{align}
	\label{eq:chi}
	\tilde{\chi}=1-\frac{M\Delta(\tilde{r})}{\tilde{r}\pa{\tilde{r}-M}^2}.
\end{align}
We will see below that $\gamma$ controls the demagnification of successive images of an isotropically emitting source, as first realized in Ref.~\cite{Johnson2019}.

It is useful to know the change in $\phi$ accrued over each orbit (period in the $\theta$-motion) of a bound photon.  This quantity was computed by Teo \cite{Teo2003}, and may also be be inferred from an $r \to \tilde{r}$ limit of the integral formulation above, as follows.  First, note from Eqs.~\eqref{eq:GeodesicIntegrals} that for $r\approx\tilde{r}$, we have
\begin{align}
	\label{eq:IphiIntegral}
	I_\phi\approx a\pa{\frac{\tilde{r}+M}{\tilde{r}-M}}I_r
	=a\pa{\frac{\tilde{r}+M}{\tilde{r}-M}}G_\theta,
\end{align}
where the last step follows from Eq.~\eqref{eq:IG}.  Letting $r\to\tilde{r}$ in Eq.~\eqref{eq:IntegralGeodesicEquation1} after using Eq.~\eqref{eq:IphiIntegral}, the change in $\phi$ for a bound photon is given in terms of angular integrals as
\begin{align}
	\Delta\phi=a\pa{\frac{\tilde{r}+M}{\tilde{r}-M}}G_\theta+\lambda G_\phi.
\end{align}
To determine the change in $\phi$ over a complete orbit, we use the formulas \eqref{eq:GthetaIntegral} and \eqref{eq:GphiIntegral} with $\theta_s=\theta_o$ and $m=2$.  Denoting this change in $\phi$ by $2\hat{\delta}$, we find
\begin{align}
	\label{eq:BoundOrbitRotation}
	\hat{\delta}=\frac{2}{a\sqrt{-\tilde{u}_-}}\br{a\pa{\frac{\tilde{r}+M}{\tilde{r}-M}}\tilde{K}+\tilde{\lambda}\tilde{\Pi}},
\end{align}
in agreement with Eq.~(18) of Ref.~\cite{Teo2003}.  This quantity $\hat{\delta}$ encodes the change in $\phi$ completed by a bound photon over each half-orbit.

As discussed in Ref.~\cite{Teo2003}, this expression for $\hat{\delta}$ is not a smooth function of $\tilde{r}$, but rather has a jump discontinuity of $2\pi$ at the pole-crossing orbit $\tilde{r}=\tilde{r}_0$.  This can be understood by imagining two photons passing nearly over the pole, but on opposite sides.  The photon moving in a locally counterclockwise direction is regarded as having accumulated approximately $\pi$ radians during the passage, whereas the clockwise photon passing on the other side is regarded as having accumulated $-\pi$ radians.  This discontinuity is essential to the mathematics of the integral formulation of the equations, but for presenting final results it will be convenient to define a continuous function by adding $2\pi$ to the $\tilde{r}>\tilde{r}_0$ branch of $\hat{\delta}$.  We will denote this smooth version by $\delta$,
\begin{align}
	\label{eq:JumpDiscontinuity}
	\delta=\hat{\delta}+2\pi H\pa{\tilde{r}-\tilde{r}_0},
\end{align}
where $H(x)$ denotes the Heaviside function.  Combining Eqs.~\eqref{eq:BoundOrbitRotation} and \eqref{eq:JumpDiscontinuity} gives
\begin{align}
	\label{eq:CriticalDelta}
	\delta=\frac{2}{\sqrt{-\tilde{u}_-}}\br{\pa{\frac{\tilde{r}+M}{\tilde{r}-M}}\tilde{K}+\frac{\tilde{\lambda}\tilde{\Pi}}{a}}+2\pi H\pa{\tilde{r}-\tilde{r}_0}.
\end{align}
We will see below that $\delta$ controls the apparent rotation of successive images of an isotropically emitting source.

Finally, consider the elapsed time $t$ over a full libration.  By a similar argument as used for Eq.~\eqref{eq:IphiIntegral}, we find
\begin{align}
	I_t\approx\tilde{r}^2\pa{\frac{\tilde{r}+3M}{\tilde{r}-M}}G_\theta.
\end{align}
Plugging this into Eq.~\eqref{eq:IntegralGeodesicEquation2} and letting $r\to\tilde{r}$ leads to
\begin{align}
	\Delta t=\tilde{r}^2\pa{\frac{\tilde{r}+3M}{\tilde{r}-M}}G_\theta+a^2G_t
\end{align}
for a bound photon orbit.  Using Eqs.~\eqref{eq:GthetaIntegral} and \eqref{eq:GtIntegral} with $\theta_s=\theta_o$ and $m=2$ gives the lapse in $t$ for a full orbit.  Denoting this time lapse over a full orbit by $2\tau$, we find
\begin{align}
	\label{eq:CriticalTau}
	\tau=\frac{2}{a\sqrt{-\tilde{u}_-}}\br{\tilde{r}^2\pa{\frac{\tilde{r}+3M}{\tilde{r}-M}}\tilde{K}-2a^2\tilde{u}_+\tilde{E}'}.
\end{align}
This quantity $\tau$ gives the change in $t$ over each half-orbit of a bound photon.  We will see below that $\tau$ controls the time-delay between the arrival of successive images of an isotropically emitting source.

\section{The screen of a distant observer}
\label{sec:Observer}

\begin{figure*}
	\includegraphics[width=\textwidth]{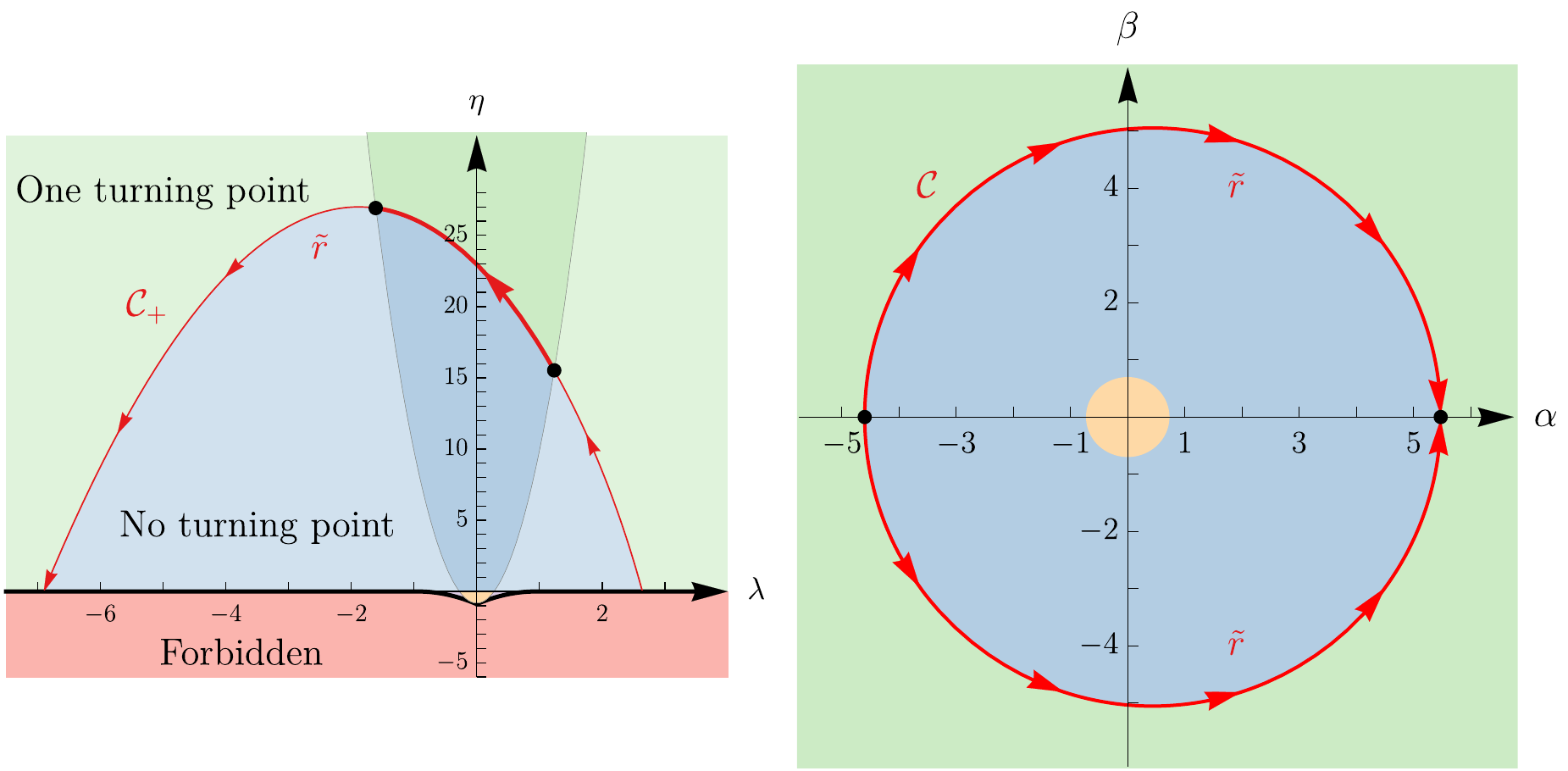}
	\caption{The 2-1 mapping from conserved quantities $(\lambda,\eta)$ to image coordinates $(\alpha,\beta)$.  The curve $\mathcal{C}_+$ of critical rays separates the regions of $(\lambda,\eta)$-space where rays have no radial turning points (blue and yellow) from the region where they have a single radial turning point (green).  (Yellow rays are vortical, while blue rays are ordinary.)  Rays can reach an observer at inclination $\theta_o$ only in the darker portion inside the gray parabola.  The 2-1 image of this portion of $\mathcal{C}_+$ defines the image-plane critical curve $\mathcal{C}$.  As $\theta_o\to0,$ the parabola closes to the vertical half-line $\lambda=0$, $\eta>-a^2$, while as $\theta_o\to\pi/2$, it opens up to a horizontal line $\eta=0$, such that the entire blue and green regions (and none of the yellow region) map to the image.  As $a\to0$, the vortical region disappears from both plots (no vortical geodesics exist).  In these plots, we chose $a/M=94\%$, $\theta_o=17^\circ$, and set $M=1$.}
	\label{fig:2-1}
\end{figure*}

Now consider a distant observer with inclination $\theta_o$ relative to the spin axis of the black hole.  We will exclude the equatorial case and use the reflection symmetry of the spacetime to place the observer in the upper hemisphere,
\begin{align}
	\theta_o\in[0,\pi/2).
\end{align}
First, consider the off-axis case $\theta_o\neq0$.  We use the axisymmetry of the spacetime to set the observer azimuthal angle to zero,
\begin{align}
	\label{eq:Observer}
	\text{Observer }\theta_o\neq0:\qquad
	r_o\to\infty,\quad
	\phi_o=0.
\end{align}

Orthogonal impact parameters $(\alpha,\beta)$ of photons reaching the observer \eqref{eq:Observer} are proportional to direction cosines on the observer's sky, and may therefore be regarded as image plane Cartesian coordinates.  Expressed in terms of photon conserved quantities, a convenient choice is \cite{Bardeen1973,Gralla2017}
\begin{align}
	\label{eq:ScreenCoordinates}
	\alpha=-\frac{\lambda}{\sin{\theta_o}},\quad
	\beta&=\pm_o\sqrt{\Theta(\theta_o)}\\
	&=\pm_o\sqrt{\eta+a^2\cos^2{\theta_o}-\lambda^2\cot^2{\theta_o}}.\nonumber
\end{align}
This defines a ``line of sight'' $\alpha=\beta=0$ to the black hole, with the $\beta$-axis regarded as the projection of the spin axis onto the plane perpendicular to this line of sight.\footnote{At large distances $r\to\infty$, the Boyer-Lindquist coordinates define a fiducial flat metric whose $z$-axis is identified with the spin axis of the black hole.  The photon with $\alpha=\beta=0$ is aimed radially inward and reaches the origin $r=0$ of this auxiliary spacetime.   The $\beta$-axis is the projection of the $z$-axis onto the ``image plane'' perpendicular to this line of sight (e.g., Fig.~6 of Ref.~\cite{Gralla2017}).}  The projected black hole rotation is in the counterclockwise direction as seen by the observer.  In comparing to an observed image, one may rescale $\alpha$ and $\beta$ to adjust for angular size, translate or rotate to adjust for the position and orientation of the source, and reverse the handedness $\alpha \to -\alpha$ to account for the projected black hole spin direction.  Finally, notice that we have
\begin{align}
	\label{eq:BetaSign}
	\pm_o=\sign\pa{\beta}.
\end{align}

Rays that reach our distant observer may have two qualitatively different origins: they either came from the white hole, or else from the celestial sphere.  Equivalently, we may imagine tracing a photon back in time from the observer and asking whether it ``ends up'' (started) at the horizon $r=r_+$, or at infinity $r\to\infty$.\footnote{One could also pose the problem forward in time, sending photons toward the black hole from the observer at infinity.  However, the black hole must then rotate in the opposite sense, as can be seen from the discrete $t\to-t$, $\phi\to-\phi$ symmetry of the metric.}  The boundary between these two behaviors corresponds to a ray that, when traced backwards in time, orbits indefinitely as it approaches a bound orbit at some radius $\tilde{r}$.  Such rays must have the same conserved quantities $\tilde{\lambda}(\tilde{r})$ and $\tilde{\eta}(\tilde{r})$ [given in Eqs.~\eqref{eq:lambdaCritical} and \eqref{eq:etaCritical} above] as the bound photon orbits.  This condition defines the critical curve $\mathcal{C}$.

The radius of the associated photon orbit provides a convenient parameterization of $\mathcal{C}$,
\begin{align}
	\label{eq:CurveCoordinates}
	\tilde{\alpha}=\alpha(\tilde{\lambda}(\tilde{r})),\quad
	\tilde{\beta}=\beta(\tilde{\lambda}(\tilde{r}),\tilde{\eta}(\tilde{r})), 
\end{align}
defined using Eqs.~\eqref{eq:lambdaCritical}, \eqref{eq:etaCritical}, and \eqref{eq:ScreenCoordinates}.  In light of the sign $\pm_o=\sign\pa{\beta}$ in Eqs.~\eqref{eq:ScreenCoordinates}, Eq.~\eqref{eq:CurveCoordinates} really refers to two separate parameterized curves (one in the upper half-plane and one in the lower half), whose union gives rise to the closed curve $\mathcal{C}$ on the image plane.  Put differently, the critical curve is a 2-1 mapping from the critical locus in conserved quantity space (Fig.~\ref{fig:2-1}).  In particular, $\mathcal{C}$ is reflection-symmetric about the $\alpha$ axis.  The range of the parameter $\tilde{r}$ is determined by the requirement that $\tilde{\beta}$ be real, which restricts to bound photon orbits for which nearby photons can escape to infinity at the observer inclination $\theta_o$ (see Fig.~2 of Ref.~\cite{Johnson2019}).  In the edge-on case $\theta_o=\pi/2$, this corresponds to the full range $\tilde{r}\in\br{\tilde{r}_-,\tilde{r}_+}$ of bound orbits in the photon shell [Eq.~\eqref{eq:PhotonShell}], whereas at smaller inclinations, there is a smaller range that can be determined numerically by finding the roots of $\tilde{\beta}(\tilde{r})$.

The shape of the critical curve depends on the black hole spin $a$ and the observer inclination $\theta_o$.  However, it is very nearly circular everywhere across this parameter space, except in the extremal, edge-on limit, where it becomes flattened on one side \cite{Bardeen1973,Falcke2000,Johannsen2010,Gralla2017,Johnson2019}.

It is useful to have a simple test of whether a given screen position $(\alpha,\beta)$ lies inside the critical curve.  One method is to compute $(\lambda,\eta)$ via the inversion of Eq.~\eqref{eq:ScreenCoordinates},
\begin{align}
	\label{eq:lambdaScreen}
	\lambda&=-\alpha\sin{\theta_o},\\
	\label{eq:etaScreen}
	\eta&=\pa{\alpha^2-a^2}\cos^2{\theta_0}+\beta^2,
\end{align}
and then plug these parameters into the formula \eqref{eq:r4} for the radial root $r_4$, which is always the outermost turning point outside the horizon (when it exists).  That is,
\begin{itemize}
	\item[] The screen point $(\alpha,\beta)$ is outside $\mathcal{C}$ if $r_4(\alpha,\beta)$ is real and outside the horizon; otherwise it lies inside $\mathcal{C}$ [$r_4$ is constructed from Eqs.~\eqref{eq:lambdaScreen}, \eqref{eq:etaScreen}, and \eqref{eq:r4}].
\end{itemize}

\subsection{On-axis observer}
\label{sec:ObserverOnAxis}

In the special case $\theta_o=0$ of an on-axis observer, it is more convenient to use polar coordinates $(b,\varphi)$ on the image plane.  Here, $b$ is the impact \textit{radius} $b=\sqrt{\alpha^2+\beta^2}$ and $\varphi$ is the angle of arrival,
\begin{align}
	\label{eq:ScreenAngle}
	\varphi=\phi_o\qquad
	\pa{\theta_0=0,\ r_o\to\infty}.
\end{align}
Since photons that reach the pole must have vanishing azimuthal angular momentum ($\lambda=0$), it follows from Eqs.~\eqref{eq:ScreenCoordinates} that
\begin{align}
	\label{eq:ScreenRadius}
	b=\sqrt{\eta+a^2}.
\end{align}
Moreover, since all photons reach a polar observer with negative $p_o^\theta$, we also have from Eq.~\eqref{eq:BetaSign} that
\begin{align}
	\label{eq:OnAxisSign}
	\pm_o=-1.
\end{align}

To simplify expressions in the case of a polar observer, we send 
\begin{align}
	\label{eq:OnAxis}
	\lambda\to0,\quad
	\eta\to b^2-a^2,
\end{align}
which in particular sends
\begin{align}
	\label{eq:u}
	u_+\to1,\quad
	u_-\to1-\frac{b^2}{a^2},
\end{align}
as well as
\begin{align}
	\theta_-\to0,\quad
	\theta_+\to\pi.
\end{align}
In most expressions, one can simply set these values, but more care is needed near turning points (pole crossings).  In particular, the angle $\phi$ jumps by $\phi\to\phi+\pi$ discontinuously at each turning point.  This coordinate artefact is reflected in the mathematics as a divergence of the angular integral $G_\phi$ at each turning point.  The relevant finite limit (recalling that $\eta>0$) is
\begin{align}
	\label{eq:PoleJump}
	\lim_{\lambda\to0^\pm}\frac{2\lambda\Pi}{a\sqrt{-u_-}}=\pm\pi.
\end{align}

The critical curve of a polar observer is a perfect circle centered at the origin.  The range of $\tilde{r}$ degenerates to a single value $\tilde{r}=\tilde{r}_0$, which is the unique radius \eqref{eq:CriticalRadius} in the photon shell $\br{\tilde{r}_-,\tilde{r}_+}$ that admits pole-crossing bound orbits ($\tilde{\lambda}=0$).  That is, from the perspective of a polar observer, the only visible portion of the photon shell is a photon sphere.  The critical curve radius $\tilde{b}=\sqrt{\tilde{\eta}+a^2}$ is given by
\begin{align}
	\label{eq:CriticalImpactParameter}
	\tilde{b}=\sqrt{\frac{\tilde{r}_0^3}{a^2}\br{\frac{4M\Delta(\tilde{r}_0)}{\pa{\tilde{r}_0-M}^2}-\tilde{r}_0}+a^2}.
\end{align}
In this case, the angle $\varphi=\phi_o$ may be viewed as the parameter along $\mathcal{C}$.

Using Eqs.~\eqref{eq:OnAxis}, \eqref{eq:u}, and \eqref{eq:PoleJump}, the critical parameters $\gamma$, $\delta$, and $\tau$ reduce to
\begin{align}
	\label{eq:gamma0}
	\gamma_0&=\frac{4\tilde{r}_0}{\sqrt{\tilde{b}^2-a^2}}\sqrt{1-\frac{M\Delta(\tilde{r}_0)}{\tilde{r}_0\pa{\tilde{r}_0-M}^2}}K\pa{\frac{a^2}{a^2-\tilde{b}^2}},\\
	\label{eq:delta0}
	\delta_0&=\pi+\frac{2a}{\sqrt{\tilde{b}^2-a^2}}\pa{\frac{\tilde{r}_0+M}{\tilde{r}_0-M}}K\pa{\frac{a^2}{a^2-\tilde{b}^2}},\\
	\label{eq:tau0}
	\tau_0&=\frac{2}{\sqrt{\tilde{b}^2-a^2}}\bigg[\tilde{r}_0^2\pa{\frac{\tilde{r}_0+3M}{\tilde{r}_0-M}}K\pa{\frac{a^2}{a^2-\tilde{b}^2}}\nonumber\\
	&\qquad\qquad\qquad-2a^2E'\pa{\frac{a^2}{a^2-\tilde{b}^2}}\bigg].
\end{align}

In the limit $a\to0$ of a nonspinning black hole (where any observer can be made polar by rotational symmetry), these quantities simplify tremendously:
\begin{align}
	\tilde{r}_0&=3M,
	&\tilde{b}=3\sqrt{3}M,\\
	\gamma_0&=\delta_0=\pi,
	&\tau_0=3\sqrt{3}\pi M.
\end{align}
These critical parameters characterize the critical orbits in the photon spheret of the Schwarzschild spacetime.

It is helpful to contrast the cases of on-axis and off-axis observers.  In the off-axis case $\theta_o\neq0$, we set the azimuthal coordinate to a fiducial value $\phi_o=0$, and the two conserved quantities $\lambda$ and $\eta$ (together with the sign $\pm_o$) encode the arrival position of photons via Eqs.~\eqref{eq:ScreenCoordinates}.  On the other hand, in the on-axis case $\theta_o=0$, one conserved quantity $\lambda$ always vanishes, and the arrival position is encoded by the second conserved quantity $\eta$ together with the azimuthal coordinate $\phi_o$ via Eqs.~\eqref{eq:ScreenAngle} and \eqref{eq:ScreenRadius}.  The critical curve has a similar shape in each case but a rather different mathematical description: for off-axis observers, we parameterize it by $\tilde{r}$, while for on-axis observers, we have $\tilde{r}=\tilde{r}_0$ and the curve is instead parameterized by $\varphi$ (and given by $b=\tilde{b}$).

\section{Behavior of rays}
\label{sec:Rays}

\begin{figure*}
	\centering
	\vspace{-10pt}
	\begin{overpic}[width=\textwidth]{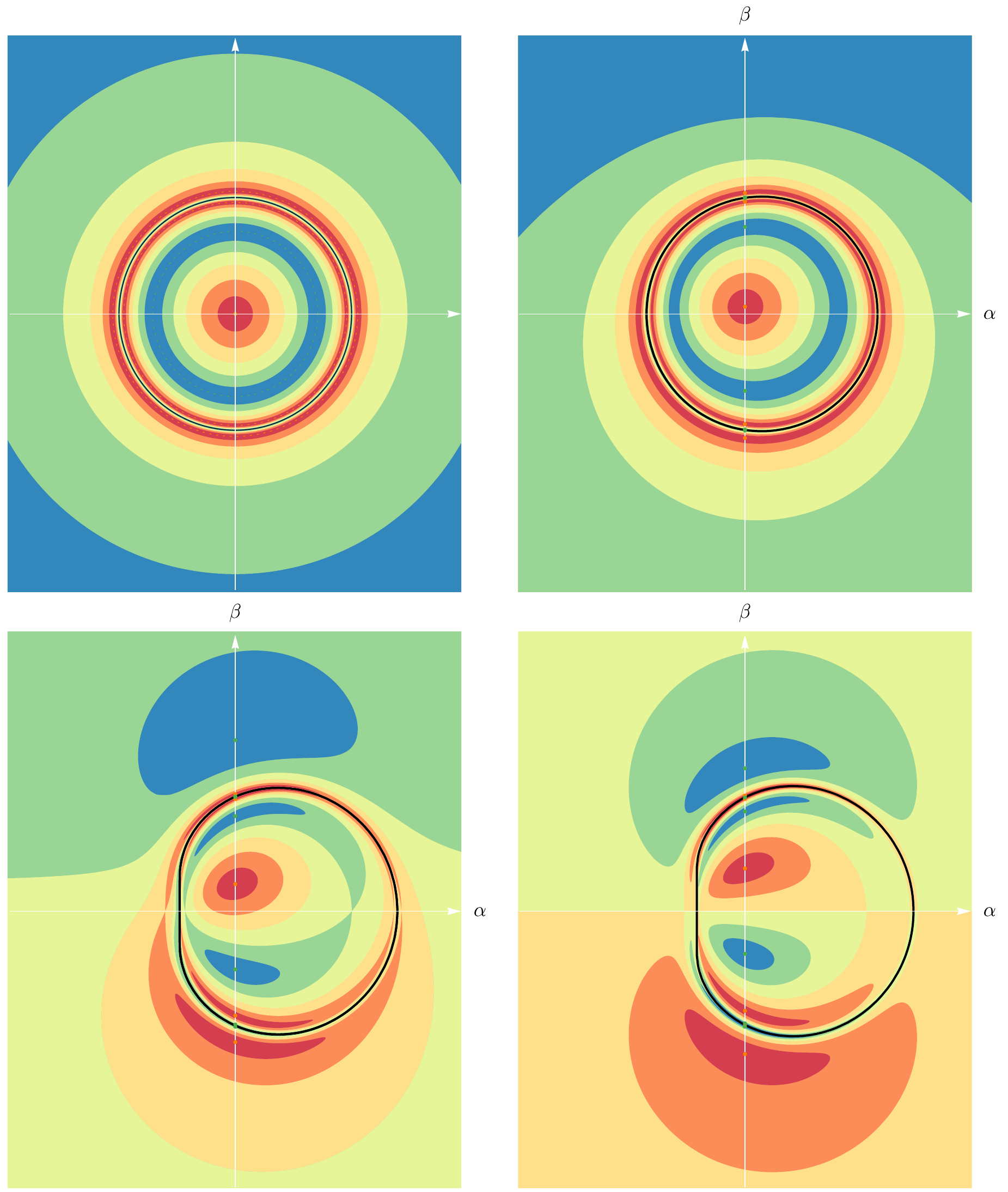}
		\put(36.9,44.4){\includegraphics[width=.1\textwidth]{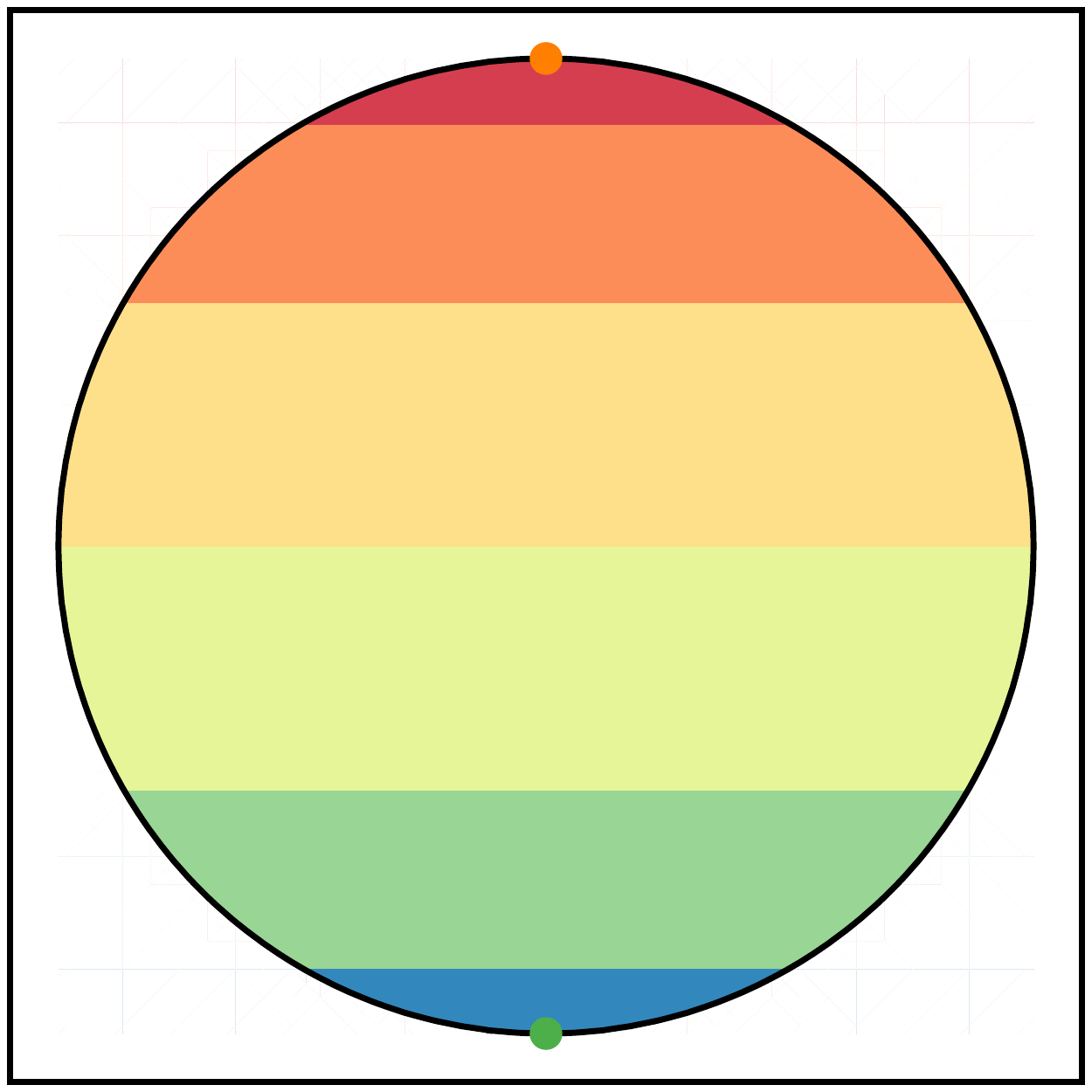}}
	\end{overpic}
	\caption{Latitude bands of the event horizon and celestial sphere, as seen by a distant observer.  Rays from the horizon (emitted just outside the black hole, or emerging from the white hole) arrive within the critical curve (black), while rays from the celestial sphere arrive outside of it.  We show the screen position of these rays, colored by the latitude of emission on the event horizon or celestial sphere, as shown in central inset (colors change every $30^{\circ}$, with orange/green dots depicting the north/south poles).  The observer sees infinitely many ``unfoldings'' of both the horizon and the celestial sphere.  Here, we show an extreme black hole ($a=M$) as viewed by a distant observer at inclinations (clockwise from top left) $\theta_o=0^\circ$, $17^\circ$, $60^\circ$, and $90^\circ$.}
	\label{fig:Unfolding}
\end{figure*}

\begin{figure*}
	\includegraphics[width=\textwidth]{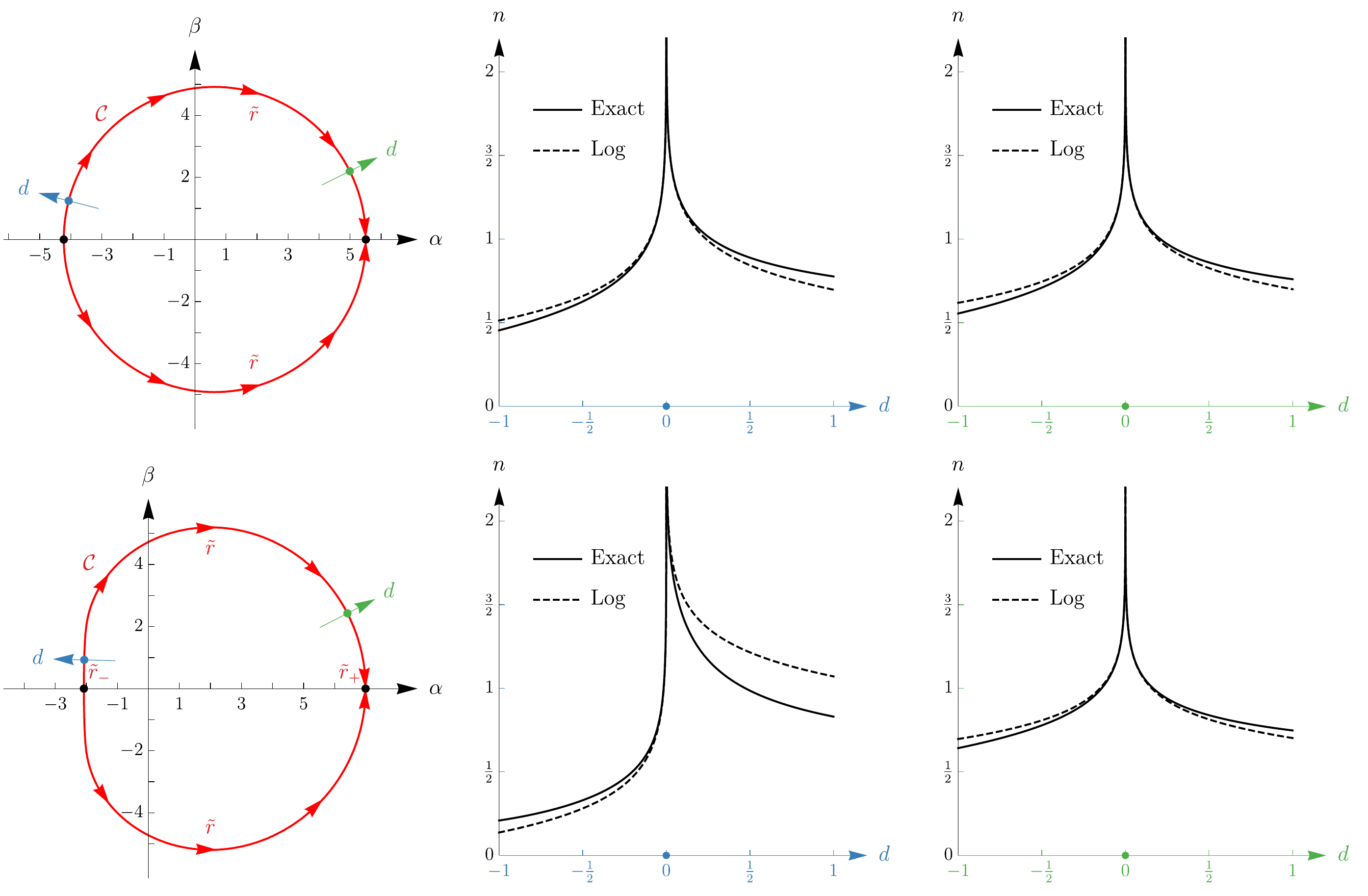}
	\caption{Fractional number of orbits $n$ as a function of signed perpendicular distance $d$ from the critical curve $\mathcal{C}$ on a distant observer's image plane.  Top: black hole spin $a/M=94$\% and observer inclination $\theta_o=17^\circ$; bottom: spin $a/M=99.9$\% and inclination $\theta_o=90^\circ$.  The curve $\mathcal{C}$ is parameterized in two separate segments above and below the $\alpha$-axis by the radius $\tilde{r}$ that rays asymptotically approach.  (The directions of increasing $\tilde{r}$ are indicated on each segment by red arrows.  The range of $\tilde{r}$ is determined by the condition $\beta^2\ge0$; only the equatorial observer $\theta_o=\pi/2$ sees the entire range $\tilde{r}\in\br{\tilde{r}_-,\tilde{r}_+}$ of bound photon orbits.)  Physically, the coordinate system $(\tilde{r},d)$ labels (nearly) bound photons by the Boyer-Lindquist radius $\tilde{r}$ of their (nearby) spherical photon orbit.  The fractional number of orbits diverges logarithmically as $\ab{d}\to0$.  The logarithmic approximation [Eq.~\eqref{eq:nCritical}] is excellent within a distance ${\sim}M$ of the critical curve (we set $M=1$ in all the plots), except near the vertical straight line (``NHEKline'') that appears in the extremal limit for $\theta_o\gtrsim47^\circ$ and requires a separate analytic treatment \cite{Gralla2017}.}
	\label{fig:Screen}
\end{figure*}

\begin{figure*}
	\centering
	\includegraphics[width=\textwidth]{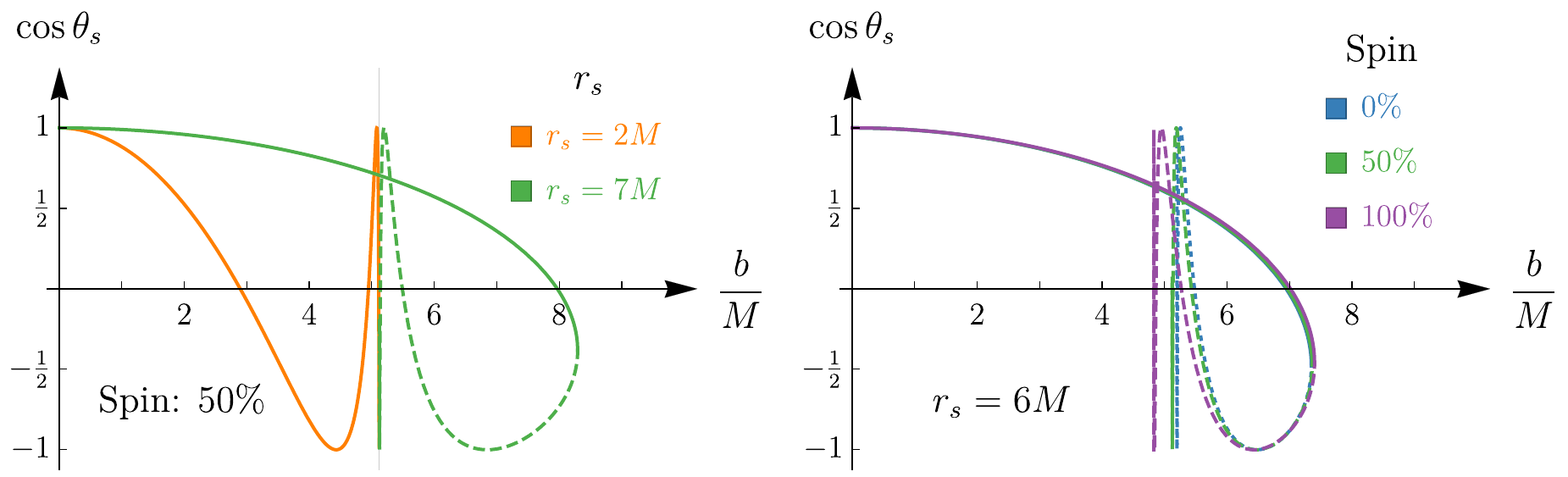}
	\caption{Behavior of photons emitted from a source sphere $r=r_s$ and received at the pole $\theta_o=0$.  We show the cosine of the emission latitude $\theta_s$ as a function of screen radius $b$.  Each oscillation from $+1$ to $-1$ represents an image of the source sphere.  When the source sphere is inside the photon sphere ($r_s<\tilde{r}_0$), the images do not overlap---the sphere is ``unwrapped'' infinitely many times on the image plane.  When the source sphere is outside the photon sphere, its first image is folded on itself, and subsequent images are superposed on this first image.  If the sphere is optically thick, emission corresponding to dashed lines will not be visible.  In the flat spacetime this corresponds to the statement that one sees only the top half of the sphere.}
	\label{fig:Sphere}
\end{figure*}

We now make some general comments about the properties of rays, i.e., complete null geodesics in the Kerr exterior.  Their radial integral $I_r$ is the total integral discussed in Eq.~\eqref{eq:IrTotal} above.  Plugging Eqs.~\eqref{eq:IG} and \eqref{eq:BetaSign} into Eq.~\eqref{eq:SourceAngle}, we find that (regardless of the sign of $\eta$)
\begin{align}
	\label{eq:OriginalAngle}
	\frac{\cos\theta_s}{\sqrt{u_+}}=\sn\pa{F_o+\sign\pa{\eta\beta}a\sqrt{-u_-}I_r^\mathrm{total}\left|\frac{u_+}{u_-}\right.}.
\end{align}
The formula \eqref{eq:OriginalAngle} gives the latitude at which the ray arriving at screen coordinate $(\alpha,\beta)$ entered the spacetime (either from the white hole if arriving inside $\mathcal{C}$, or from the celestial sphere if arriving outside $\mathcal{C}$).  The level sets of this function show how the horizon and celestial sphere are ``unfolded'' infinitely many times on the image plane, converging to the critical curve (Fig.~\ref{fig:Unfolding}).

Each successive unfolding corresponds to a photon that has undergone an additional half-orbit before reaching the observer.  To study this effect quantitatively, we consider the total (fractional) number of oribts $n$, which is proportional to $I_r$ by Eq.~\eqref{eq:n}.  The results of App.~\ref{app:MAE} provide an asymptotic expansion valid for near-critical rays.  From Eqs.~\eqref{eq:n}, \eqref{eq:Outside}, \eqref{eq:Inside} and \eqref{eq:Distance}, we have 
\begin{align}
	\label{eq:nCritical}
	n\approx-\frac{1}{2\gamma(\tilde{r})}\log\br{\hat{C}_\pm(\tilde{r})d},\quad
	d\to0^\pm, 
\end{align}
where $d$ is the signed perpendicular distance from the closest point $\tilde{r}$ on the critical curve, $\gamma(\tilde{r})$ is the Lyapunov exponent \eqref{eq:LyapunovExponent}, and we also introduced coefficients
\begin{align}
	\label{eq:OutsideCoefficient}
	\hat{C}_+(\tilde{r})&=\pa{\frac{1+\sqrt{\tilde{\chi}}}{8\tilde{\chi}}}^2\frac{\Delta(\tilde{r})}{2\tilde{r}^4\tilde{\chi}}\sqrt{\tilde{\beta}^2+\tilde{\psi}^2},\\
	\label{eq:InsideCoefficient}
	\hat{C}_-(\tilde{r})&=-\frac{\sqrt{1-\tilde{\chi}}}{1+\sqrt{\tilde{\chi}}}\sqrt{\frac{1+\mathcal{Q}_2(\dt r_+,0)}{1-\mathcal{Q}_2(\dt r_+,0)}}\hat{C}_+(\tilde{r}).
\end{align}
See Eqs.~\eqref{eq:chi}, \eqref{eq:CurveCoordinates}, \eqref{eq:dtrpm}, \eqref{eq:Q2} and \eqref{eq:psiCritical} for definitions of the various quantities that appear.   In the nonrotating limit $a\to0$, Eqs.~\eqref{eq:nCritical}, \eqref{eq:OutsideCoefficient} and \eqref{eq:InsideCoefficient} agree with Eqs.~(2), (3), and (4) of Ref.~\cite{Gralla2019}.  The exact and approximate fractional number of orbits are shown in Fig.~\ref{fig:Screen}.

As depicted in Fig.~\ref{fig:Screen}, we may think of $(\tilde{r},d)$ as a set of coordinates for the image plane that are defined in the neighborhood of $\mathcal{C}$ for which there is a unique line segment connecting any point $p$ to $\mathcal{C}$, with the line intersecting $\mathcal{C}$ perpendicularly.  The coordinate $\tilde{r}$ of the point $p$ is the Boyer-Lindquist radius of the associated photon orbit where $\mathcal{C}$ is intersected, and the coordinate $d$ is the signed length of the segment (i.e., $|d|$ is the length, with $d$ positive/negative when the point $p$ is outside/inside $\mathcal{C}$).  This actually defines two coordinate charts---one in the upper half-plane and one in the lower half-plane---since each radius $\tilde{r}$ corresponds to two points on $\mathcal{C}$ related by $\beta\to-\beta$.  That is, points near $\mathcal{C}$ are uniquely described by $(\tilde{r},d,\sign\pa{\beta})$.  We will generally leave the $\sign\pa{\beta}$-dependence implicit, regarding $(\tilde{r},d)$ as a single chart.  In the case of an on-axis observer $\theta_o=0$, for whom the $\tilde{r}$-parameterization breaks down, we would instead use $(\varphi,d)$, where $\varphi=\phi_o$ and $d=b-\tilde{b}$, with $\tilde{b}$ given by Eq.~\eqref{eq:CriticalImpactParameter}.

The formula \eqref{eq:nCritical} may be compared with Eq.~(11) of Ref.~\cite{Johnson2019}.  Accounting for a factor of two difference in the definition of $n$, the prefactors agree exactly, but the argument in the log differs in two ways.  First, we include the coefficients $\hat{C}_\pm$ associated with a definite physical quantity, the total (fractional) number of orbits outside the horizon.  Strictly speaking, these are subleading to the dominant $\log{d}$ term, but nonetheless they are necessary to attain any reasonable degree of accuracy.  The second difference is that the dependence on the deviation from the critical curve appears as the normal distance $d$ in place of the unspecified displacement $\delta\rho/\rho_c$ in Ref.~\cite{Johnson2019}, making precise the scaling argument given therein.

\section{Behavior of photons}
\label{sec:Photons}

\begin{figure*}
	\centering
	\includegraphics[width=\textwidth]{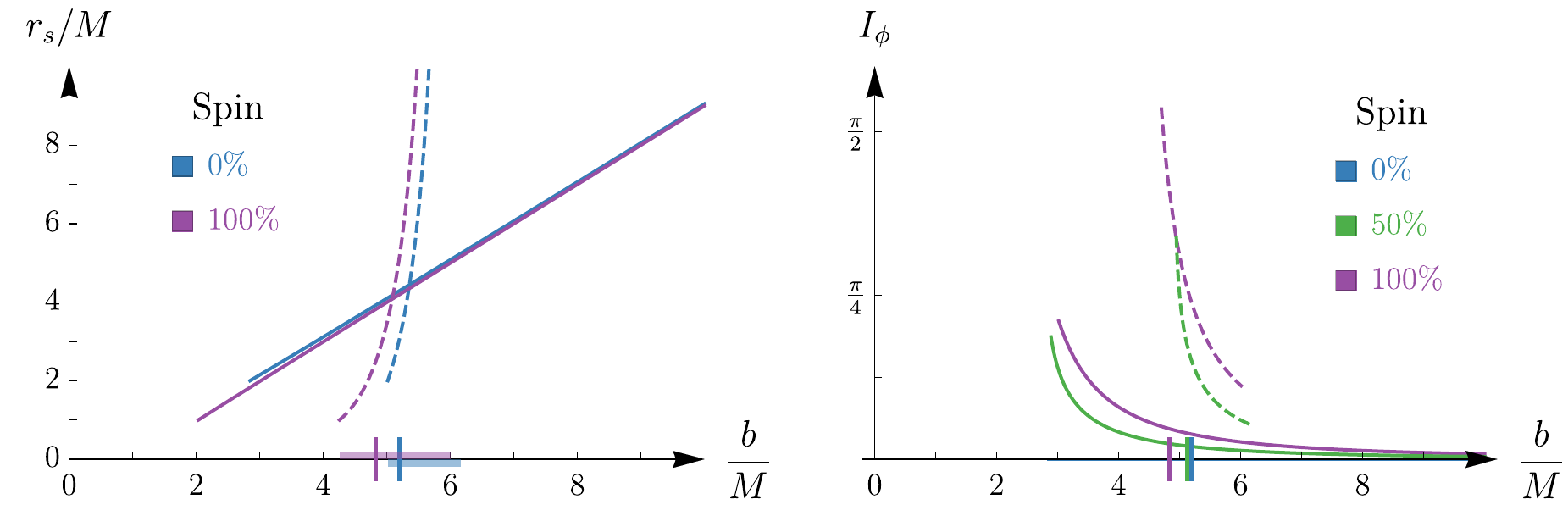}
	\caption{Behavior of photons emitted from the equatorial plane $\theta_s=\pi/2$ and received at the pole $\theta_o=0$.  Solid lines correspond to ``direct'' photons with no angular turning points ($m=0$), while dashed lines correspond to ``backward-emitted'' photons that bend around the black hole before reaching the observer ($m=1$).  The color bands on the horizontal axis show the range over which the backward-emitted photons can reach the observer [the apparent $m=1$ range of $r_s\in(+,\infty)$], and the colored ticks represent the critical curve radius $\tilde{b}$.  Higher-order photons ($m\geq2$) produce essentially vertical lines at the critical radius (e.g., Fig.~4 of Ref.~\cite{Gralla2019}) and are not shown here.  On the left, we show the emission radius $r_s$ as a function of screen radius $b$.  On the right, we show the frame dragging integral $I_\phi$, with the curves cut off at the apparent position of the ergosphere, where time-delay effects become essential (see further discussion in the main text).}
	\label{fig:Transfer}
\end{figure*}

We now make some general comments about the behavior of photons reaching the observer, i.e., portions of null geodesics corresponding to emission and observation of light.  We will consider the apparent positions (location on the observer screen) of various simple geometric sources.  A given source has infinitely many apparent positions (arising from photons making arbitrarily many orbits around the black hole), but throughout this section, we confine our attention to the first one or two, deferring discussion of higher-order images to Sec.~\ref{sec:PhotonRing} below.   We use the term ``position'' even when discussing extended sources; for example, the apparent positions of a source ring $(r_s,\theta_s)$ are closed curves on the image plane.

\subsection{Spheres observed from the pole}

We begin by discussing the apparent positions of latitude lines on a sphere of some radius $r_s$, as viewed from above ($\theta_o=0$).  Recall from Sec.~\ref{sec:ObserverOnAxis} that we use polar coordinates $(b,\varphi)$ on the image plane for such an observer.  Using Eqs.~\eqref{eq:IG}, \eqref{eq:OnAxis} and \eqref{eq:u}, Eq.~\eqref{eq:SourceAngle} becomes
\begin{align}
	\cos{\theta_s}=\cd\pa{\sqrt{b^2-a^2}I_r\left|\frac{a^2}{a^2-b^2}\right.},
\end{align}
where cd is the Jacobi elliptic function $\cd(\varphi|k)$.  The integral $I_r$ may be computed either numerically, or using elliptic integrals; we use expressions given in Ref.~\cite{KerrGeodesics}.  For fixed $r_s$, the radial integral $I_r$ is a function of $b$ that is single-valued for $b<\tilde{b}$ and double-valued for $b>\tilde{b}$ [see Eq.~\eqref{eq:IrIntegral}].  Thus, for $b<\tilde{b}$ there is a unique emission latitude $\theta_s$ for each radius $b$, whereas for $b>\tilde{b}$ there are two, corresponding to outward and inward emission [see Eq.~\eqref{eq:w}].  (The emission from these different points on the sphere would be superposed if the sphere is optically thin.  In flat spacetime, this would be tantamount to looking straight down through a sphere.)  The emission latitude(s) as a function of $b$ are shown in Fig.~\ref{fig:Sphere} for a selection of sphere radii $r_s$ and black hole spins $a$.

\subsection{Equatorial plane observed from the pole}

We now consider the apparent positions of rings lying on the equatorial plane ($\theta_s=\pi/2$) and observed from directly above ($\theta_o=0$).  Using Eqs.~\eqref{eq:OnAxisSign}, \eqref{eq:OnAxis}, and \eqref{eq:u}, Eq.~\eqref{eq:GthetaIntegral} becomes\footnote{The formula \eqref{eq:AngularIntegralOnAxis} holds only for $b^2>a^2$ on account of our assumption that $\eta>0$ (excluding vortical geodesics). However, vortical geodesics with $b^2<a^2$ cannot cross the equatorial plane, so there is no loss of generality for the equatorial sources that we treat here.}
\begin{align}
	\label{eq:AngularIntegralOnAxis}
	G_\theta=\frac{2m+1}{\sqrt{b^2-a^2}}K\pa{\frac{a^2}{a^2-b^2}}
	=I_r,
\end{align}
where the second equality follows from the geodesic equation $I_r=G_\theta$ [Eq.~\eqref{eq:IG}].  The condition $0<I_r<I_r^\mathrm{total}$ [Eq.~\eqref{eq:IrRange}] is thus
\begin{align}
	\label{eq:mRange}
	0<\frac{2m+1}{\sqrt{b^2-a^2}}K\pa{\frac{a^2}{a^2-b^2}}
	<I_r^\mathrm{total}.
\end{align}
This condition provides the range of integers $m$ for which there exist trajectories linking the equator and the polar observer with $m$ turning points, as a function of the image radius $b$.  For most values of $b$, only $m=0$ is allowed, with higher-order values of $m$ becoming allowed near the critical curve $\tilde{b}$, where $I_r$ diverges logarithmically.  For any value of $m\in\cu{0,1,2,\ldots}$ satisfying the condition \eqref{eq:mRange}, Eq.~\eqref{eq:SourceRadius} for $r_s(I_r)$ with Eq.~\eqref{eq:AngularIntegralOnAxis} for $I_r(b,m)$ provides the emission radius $r_s(b,m)$.  These maps $r_s(b,m)$ for $m\in\mathbb{N}$ were called ``transfer functions'' in Ref.~\cite{Gralla2019}.  In Fig.~\ref{fig:Transfer}, we show the first ($m=0$) and second ($m=1$) transfer functions, which correspond to the main images of the front and the back of an equatorial disk, respectively.  As discussed in Ref.~\cite{Gralla2019}, the ``backside image''  ($m=1$) is highly demagnified, appearing only in a thin band near the critical curve.\footnote{Ref.~\cite{Gralla2019} used the terminology ``lensing ring'' for this $m=1$ backside image, reserving ``photon ring'' for higher-order images $m\geq 2$.  Here, we include the $m=1$ image as part of the ``photon ring''.}  Subsequent (further demagnified) images will be discussed in Sec.~\ref{sec:PhotonRing} below.

The angle of arrival $\varphi$ of a photon is given by Eqs.~\eqref{eq:IntegralGeodesicEquation1}, \eqref{eq:GphiIntegral}, \eqref{eq:ScreenAngle} and \eqref{eq:PoleJump} as\footnote{Note that $\lambda\Pi_o$ is zero in this limit since the photon does not cross the pole at large $r_o$ near the far observer at $r_o\to\infty$.}
\begin{align}
	\label{eq:ScreenAngle2}
	\varphi=\phi_s+I_\phi+m \pi,
\end{align}
where we absorb the $\pm$ from Eq.~\eqref{eq:PoleJump} using $\varphi\sim\varphi+2\pi$.  The last term reflects the $m$ passages of the photon through the pole before it reaches the observer.  In the zero-spin limit, the middle term vanishes, showing that successive images of a single source appear on alternating, opposite sides of the image plane.  The middle term $I_\phi$ introduces an additional, spin-dependent shift in image plane angle $\varphi$, which we regard as the effect of frame dragging.

In Fig.~\ref{fig:Transfer}, we plot $I_\phi$ for the front side ($m=0$) and backside ($m=1$) images for a selection of spins.  For a static disk of emission with a nonaxisymmetric profile, the observed images will be rotated by this $b$-dependent factor; for example, a ``color wheel'' will appear ``swirled''.  However, a static disk cannot exist inside the ergoradius $r=2M$ (where rotation is inevitable), and we have therefore chosen to cut off the curves at the associated apparent radius $b$.  If the curves were continued inside, they would display a divergence at the apparent position of the event horizon due to the irregularity of the coordinate $\phi$.  In a physical model, time-delay effects would compensate this divergence ($\Delta t$ diverges as well) to give a regular appearance to the source.

\subsection{Equatorial plane: Inclined observer}

\begin{figure*}
	\centering
	\includegraphics[width=\textwidth]{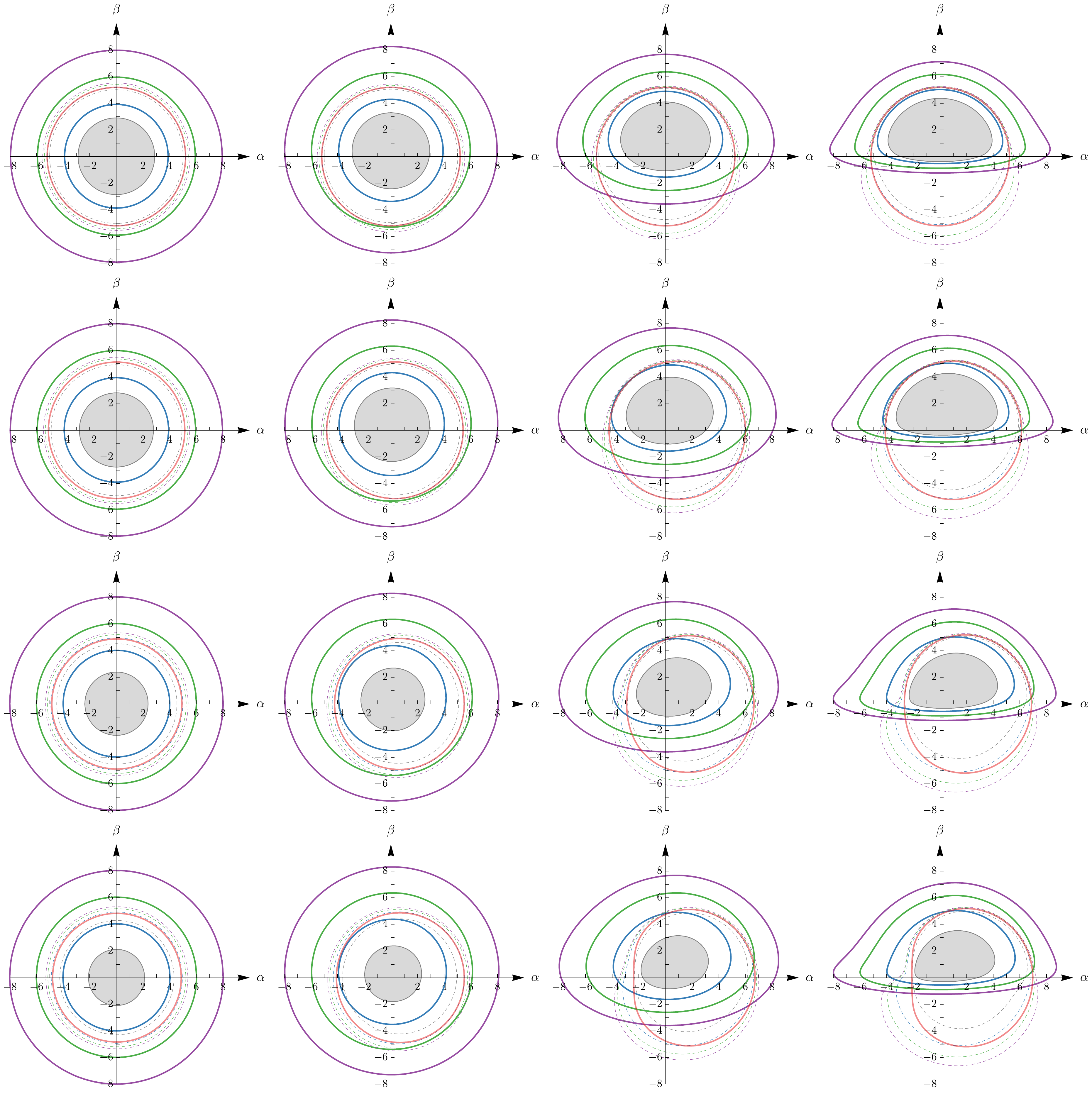}
	\caption{Apparent positions of source rings of constant Boyer-Lindquist radius $r_s$ in the equatorial plane $\theta_s=\pi/2$, as a function of black hole spin and observer inclination.  (We set $M=1$.)  Solid lines are the front side image $\bar{m}=0$, while dashed lines are the backside image $\bar{m}=1$ (Fig.~\ref{fig:Backside}).  The apparent position of the horizon is a filled gray line, while the apparent positions of $r_s=3$, 5, and 7 are blue, green, and purple, respectively.  From top to bottom, the rows are spin $a/M=1$\%, 50\%, 94\%, 99.9\%; from left to right, the columns are observer inclination $\theta_o=1^\circ$, $17^\circ$, $60^\circ$, $80^\circ$.}
	\label{fig:Contours}
\end{figure*}

\begin{figure*}
	\centering
	\includegraphics[width=\textwidth]{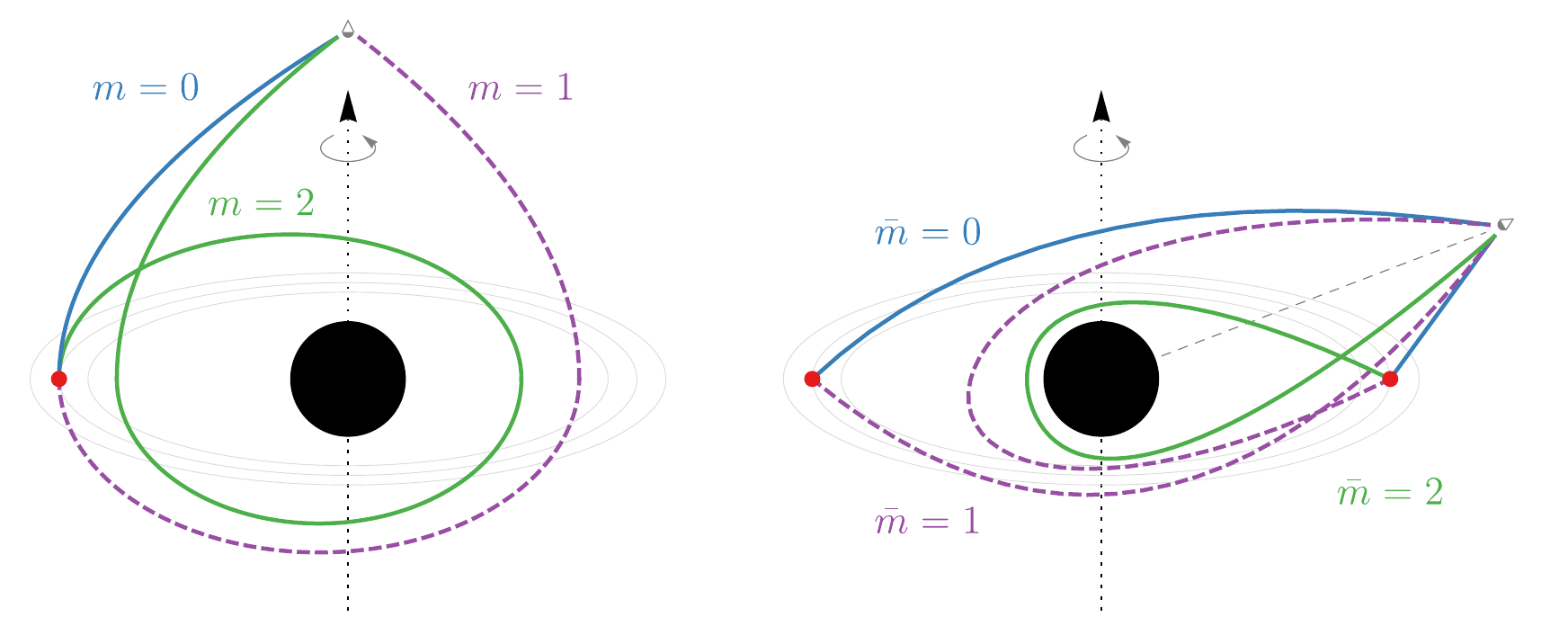}
	\caption{Illustration of the meaning of $m$ and $\bar{m}$ in the case of equatorial sources (``the disk'').  For a polar observer (left), even/odd values of $m$ correspond to emission from the front/back of the disk, and arrive on opposite sides of the image.  For the inclined observer (right), we instead use $\bar{m}$ [Eq.~\eqref{eq:mprime}], and again even/odd values come from the front/back of the disk.  Solid lines are front side images, while dashed lines are backside images.  For the left source on the right figure, we omit the $\bar{m}=2$ front side image (green) for clarity.  These curves are schematic and do not represent actual trajectories.}
	\label{fig:Backside}
\end{figure*}

We now consider equatorial sources ($\theta_s=\pi/2$) seen by inclined observers ($\theta_o\neq0$).  Noting that $F_s=0$ and $\pm_o=\sign\pa{\beta}$, Eqs.~\eqref{eq:IG} and \eqref{eq:GthetaIntegral} become 
\begin{align}
	\label{eq:EquatorialContours}
	\sqrt{-u_-a^2}I_r+\sign\pa{\beta}F_o=2mK.
\end{align}
For each $r_s$ and $m$, this equation defines a relationship between $\alpha$ and $\beta$, i.e., a curve on the image plane.  However, if this curve intersects the $\alpha$-axis, then it will be discontinuous there on account of the $\sign\pa{\beta}$ appearing in Eq.~\eqref{eq:EquatorialContours}.\footnote{If $\theta_o=\pi/2$ exactly, then $F_o$ vanishes and this jump does not occur.  We have excluded this degenerate case for simplicity.}  This jump can be simply compensated by sending $m\to m+1$ whenever the $\alpha$-axis is crossed from below, since the incomplete elliptic integral $F_o$ becomes the complete elliptic integral $K$ at $\beta=0$.\footnote{A photon arriving on the $\alpha$-axis has vanishing $\beta=p_o^\theta$, and is therefore at an angular turning point $\theta_\pm$ when it reaches the observer.}  That is, smooth curves on the image plane are labeled by integers $\bar{m}$ defined using the Heaviside function $H(x)$ by
\begin{align}
	\label{eq:mprime}
	\bar{m}=m-H(\beta).
\end{align}
This reflects the geometric fact that, since the observer is assumed to lie above the equatorial plane, emission arriving from above the line of sight must have an additional angular turning point relative to the corresponding emission arriving from below (Fig.~\ref{fig:Backside}).

Each source ring $r_s$ maps to an infinite number of observed rings labeled by $\bar{m}\in\cu{0,1,2,\dots}$.  Even $\bar{m}$ corresponds to emission towards the observer (i.e., from the front of an equatorial disk), while odd $\bar{m}$ corresponds to emission away from the observer (i.e., from the back of a disk).  In Fig.~\ref{fig:Contours}, we show the first ($\bar{m}=0$) and second ($\bar{m}=1$) rings in the form of equatorial contour plots for various values of black hole spin and inclination.  Subsequent rings ($\bar{m}\geq 2$) appear very near the critical curve and are discussed in Sec.~\ref{sec:PhotonRing} below.

\section{The photon ring}
\label{sec:PhotonRing}

We now discuss universal properties of photons arriving near the critical curve $\mathcal{C}$.  Our discussion will be framed in terms of the three key quantities $\gamma$, $\delta$, and $\tau$ that characterize the critical orbits (Sec.~\ref{sec:CriticalRays} above).  We will first derive expressions of the form
\begin{align}
	d&\propto e^{-2n\gamma},\\
	\Delta \phi&=2n\delta+(\text{corrections}),\\
	\Delta t&=2n\tau+(\text{corrections}),
\end{align}
where $n$ is the fractional number of orbits (Sec.~\ref{sec:Orbits}) and $d$ is the signed perpendicular distance from the critical curve (Fig.~\ref{fig:Screen}).  These formulas help make conceptual points about how the critical parameters $\cu{\gamma,\delta,\tau}$ of bound photon orbits influence image plane observables, but for quantitative claims, it it necessary to relate to the turning point number $m$ and discuss the corrections in detail.  For these purposes, it will be helpful to introduce the notation 
\begin{align}
	\label{eq:GeometricCorrections}
	f_i=\frac{\tilde{F}_i}{\tilde{K}},\quad
	\pi_i=\frac{\tilde{\Pi}_i}{\tilde{\Pi}},\quad
	e'_i=\frac{\tilde{E}_i}{\tilde{E}'},
\end{align}
where as usual, $i\in\cu{s,o}$ stands for source or observer.  These quantities range between $-1$ and $+1$ at $\theta_+$ and $\theta_-$, respectively, while vanishing at the equator $\theta_i=\pi/2$.

\subsection{Distance from critical curve \texorpdfstring{($\gamma$)}{}}

The analysis in App.~\ref{app:MAE} shows that near criticality, the radial integral $I_r$ evaluated from $r_s$ to $r_o\to\infty$ grows as
\begin{align}
	\label{eq:IrAsymptotic}
	I_r\approx-\frac{1}{2 \tilde{r}\sqrt{ \tilde{\chi}}} \log \left[C_\pm(r_s,\tilde{r}) d \right],
\end{align}
where $\tilde{\chi}$ is given in Eq.~\eqref{eq:chi}, while $C_\pm(r_s,\tilde{r})$ (with $\pm$ the sign of $d$) can be inferred from the equations in Secs.~\ref{sec:Outside} and \ref{sec:Inside} together with the expression \eqref{eq:Distance} for $d$.\footnote{The product $C_\pm d$ is always positive.  Note that $C_+$ is a double-valued function of $r_s$ for $d>0$, corresponding to motion before and after the turning point is reached.}

Using the geodesic equation $I_r=G_\theta$ [Eq.~\eqref{eq:IG}] in Eq.~\eqref{eq:IrAsymptotic} and solving for $d$, we obtain
\begin{align}
	\label{eq:dAsymptotic}
	d\approx\frac{1}{C_\pm(r_s,\tilde{r})}\exp\br{-2\tilde{r}\sqrt{\tilde{\chi}}G_\theta(m,\theta_s,\theta_o)}.
\end{align}
A more illuminating form of this equation is
\begin{align}
	\label{eq:dLyapunov}
	d\approx\frac{1}{C_\pm}e^{-2n\gamma},
\end{align}
where $n$ is the fractional number of orbits [Eq.~\eqref{eq:n}] and $\gamma(\tilde{r})$ is the Lyapunov exponent of the photon orbit at radius $\tilde{r}$ [Eq.~\eqref{eq:LyapunovExponent}].  Thus, for each factor $e^{-2\gamma}$ closer to the critical curve, the observed photon has executed one additional orbit.  We may relate $n=G_\theta/G_1$ to the number of polar turning points $m$ by using Eqs.~\eqref{eq:GthetaIntegral} and \eqref{eq:G1}, and setting the conserved quantities equal to their critical values,
\begin{align}
	\label{eq:OrbitNumberTurningPoint}
	n\approx\frac{m}{2}\pm_o\frac{1}{4}\br{(-1)^mf_s-f_o},
\end{align}
where the geometric factor $f_i$ was introduced in Eq.~\eqref{eq:GeometricCorrections}.

Eqs.~\eqref{eq:dAsymptotic} and \eqref{eq:dLyapunov} are valid for $d \ll M$, or equivalently for $n\gg1$ or $m\gg1$.  In practice, we find that the agreement is reasonable even for $d\sim M$ (Fig.~\ref{fig:Screen}), and hence for $n\sim1$.  In particular, the logarithmic approximation is already useful at $m=1$, and it becomes excellent for all higher $m\in\cu{2,3,4,\ldots}$.

For each value of $\tilde{r}$, around the the curve $\mathcal{C}$, and for each choice of integer $m$ (typically accurate for $m\gtrsim1$), Eq.~\eqref{eq:dLyapunov} provides the signed perpendicular distance $d$ of an arriving photon that originated on the poloidal ring $(r_s,\theta_s)$ and encountered $m$ angular turning points on its way to the observer.  The emission angle along the ring, as well as the emission time, may be found from $\Delta\phi$ and $\Delta t$, which we now discuss.

\subsection{Lapse in azimuthal angle \texorpdfstring{($\delta$)}{}}

Now, consider the lapse in $\phi$ [Eq.~\eqref{eq:IntegralGeodesicEquation1}],
\begin{align}
	\label{eq:phiLapse}
	\Delta\phi=I_\phi+\lambda G_\phi(m,\theta_s,\theta_o).
\end{align}
The analysis of App.~\ref{app:MAE} shows that near criticality, the integral $I_\phi$ takes the asymptotic form
\begin{align}
	\label{eq:IphiAsymptotic}
	I_\phi\approx a\pa{\frac{\tilde{r}+M}{\tilde{r}-M}}I_r+D_\pm(\tilde{r},r_s),
\end{align}
where the precise form of $D_\pm(\tilde{r})$ may be inferred from the expressions in App.~\ref{app:MAE}.  For our present purposes, the only important property of $D_\pm(\tilde{r})$ is that it is independent of $d$, except via the sign $\pm=\sign\pa{d}$.  Using the geodesic equation $I_r=G_\theta$ [Eq.~\eqref{eq:IG}], Eqs.~\eqref{eq:phiLapse} and \eqref{eq:IphiAsymptotic} give
\begin{align}
	\Delta\phi\approx a\pa{\frac{\tilde{r}+M}{\tilde{r}-M}}G_\theta+\tilde{\lambda}G_\phi+D_\pm(\tilde{r}),
\end{align}
such that the $d$-dependence drops out entirely, other than via $\pm=\sign\pa{d}$.   A more illuminating form of this expression is [combining Eqs.~\eqref{eq:GthetaIntegral}, \eqref{eq:GphiIntegral}, \eqref{eq:BoundOrbitRotation}, and \eqref{eq:OrbitNumberTurningPoint}] 
\begin{align}
	\label{eq:phiAsymptotic}
	\Delta\phi\approx2n\hat{\delta}-J_\pm^\phi(m,\tilde{r}),
\end{align}
where $2\hat{\delta}(\tilde{r})$ is the lapse in $\phi$ per orbit of a bound photon at radius $\tilde{r}$ [Eq.~\eqref{eq:CriticalDelta}], and
\begin{align}
	J_\pm^\phi=\frac{\pm_o\tilde{\lambda}\tilde{\Pi}}{a\sqrt{-\tilde{u}_-}}\br{(-1)^m\pa{f_s-\pi_s}-\pa{f_o-\pi_o}}-D_\pm(\tilde{r}).
\end{align}
Once again, we remind the reader that here, the subscript $\pm$ is the sign of $d$, encoding whether one is inside $(-)$ or outside $(+)$ the critical curve.

\subsection{Lapse in time \texorpdfstring{($\tau$)}{}}

Finally, consider the lapse in $t$ [Eq.~\eqref{eq:IntegralGeodesicEquation2}],
\begin{align}
	\label{eq:tLapse}
	\Delta t=I_t+a^2G_t(m,\theta_s,\theta_o).
\end{align}
The analysis of App.~\ref{app:MAE} shows that near criticality, the integral $I_t$ takes the asymptotic form
\begin{align}
	\label{eq:ItAsymptotic}
	I_t\approx\tilde{r}^2\pa{\frac{\tilde{r}+3M}{\tilde{r}-M}}I_r+H_\pm(\tilde{r},r_s),
\end{align}
where the precise form of $H_\pm(\tilde{r})$ may be inferred from the expressions in App.~\ref{app:MAE}.  For our present purposes, the only important property of $H_\pm(\tilde{r})$ is once again that it is independent of $d$, except via its sign $\pm=\sign\pa{d}$.  Using the geodesic equation $I_r=G_\theta$ [Eq.~\eqref{eq:IG}], Eqs.~\eqref{eq:tLapse} and \eqref{eq:ItAsymptotic} become
\begin{align}
	\Delta t\approx\tilde{r}^2\pa{\frac{\tilde{r}+3M}{\tilde{r}-M}}G_\theta+a^2G_t+H_\pm(\tilde{r}),
\end{align}
such that the $d$-dependence drops out entirely, other than via $\pm=\sign\pa{d}$.  A more illuminating form of this expression is [combining Eqs.~\eqref{eq:GthetaIntegral}, \eqref{eq:GtIntegral}, \eqref{eq:CriticalTau}, and \eqref{eq:OrbitNumberTurningPoint}]
\begin{align}
	\label{eq:tAsymptotic}
	\Delta t\approx2n\tau-J_\pm^t(m,\tilde{r}),
\end{align}
where $2\tau(\tilde{r})$ is the lapse in $t$ per orbit of a bound photon at radius $\tilde{r}$ [Eq.~\eqref{eq:CriticalTau}], and
\begin{align}
	J_\pm^t=\frac{\mp_o2a\tilde{u}_+ \tilde{E}'}{\sqrt{-\tilde{u}_-}}\br{(-1)^m\pa{f_s-e'_s}-\pa{f_o-e'_o}}-H_\pm(\tilde{r}).
\end{align}
Yet again, we remind the reader that here, the subscript $\pm$ is the sign of $d$, encoding whether one is inside $(-)$ or outside $(+)$ the critical curve.

\begin{figure}
	\centering
	\includegraphics[width=\columnwidth]{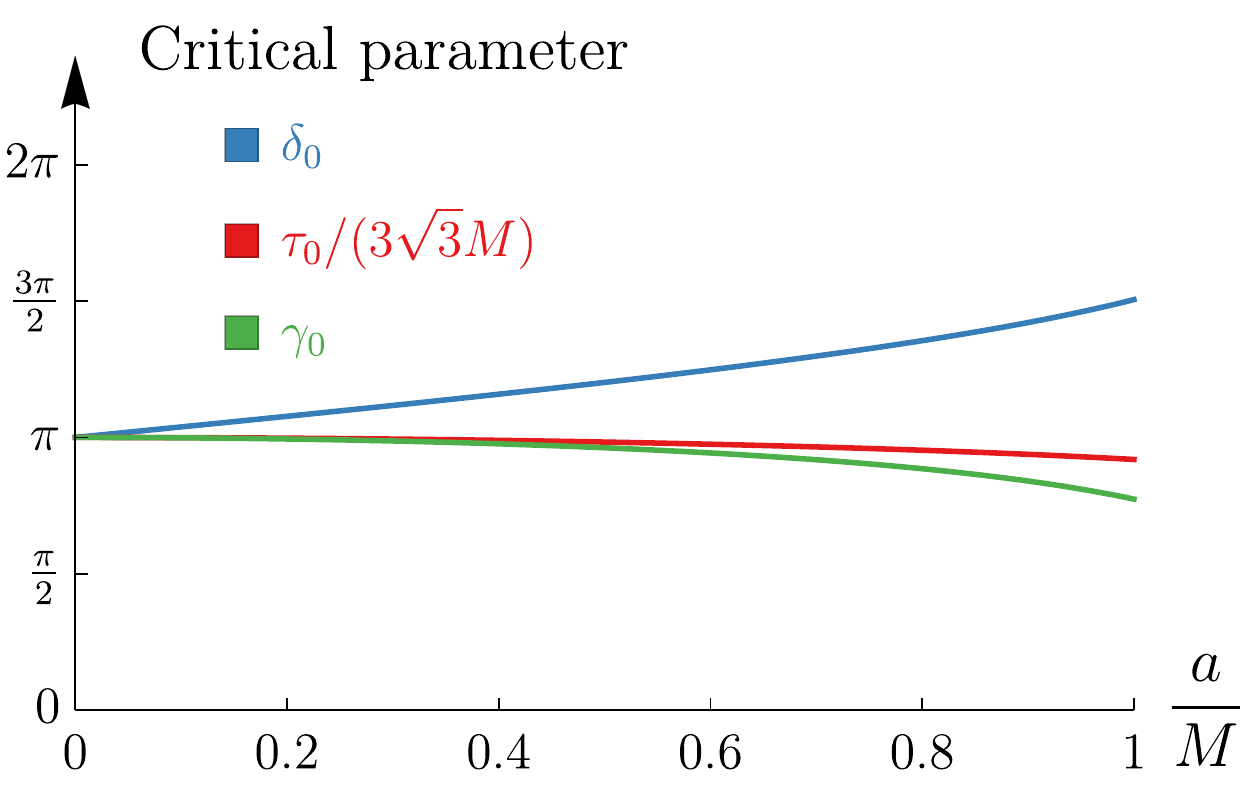}\\
	\includegraphics[width=\columnwidth]{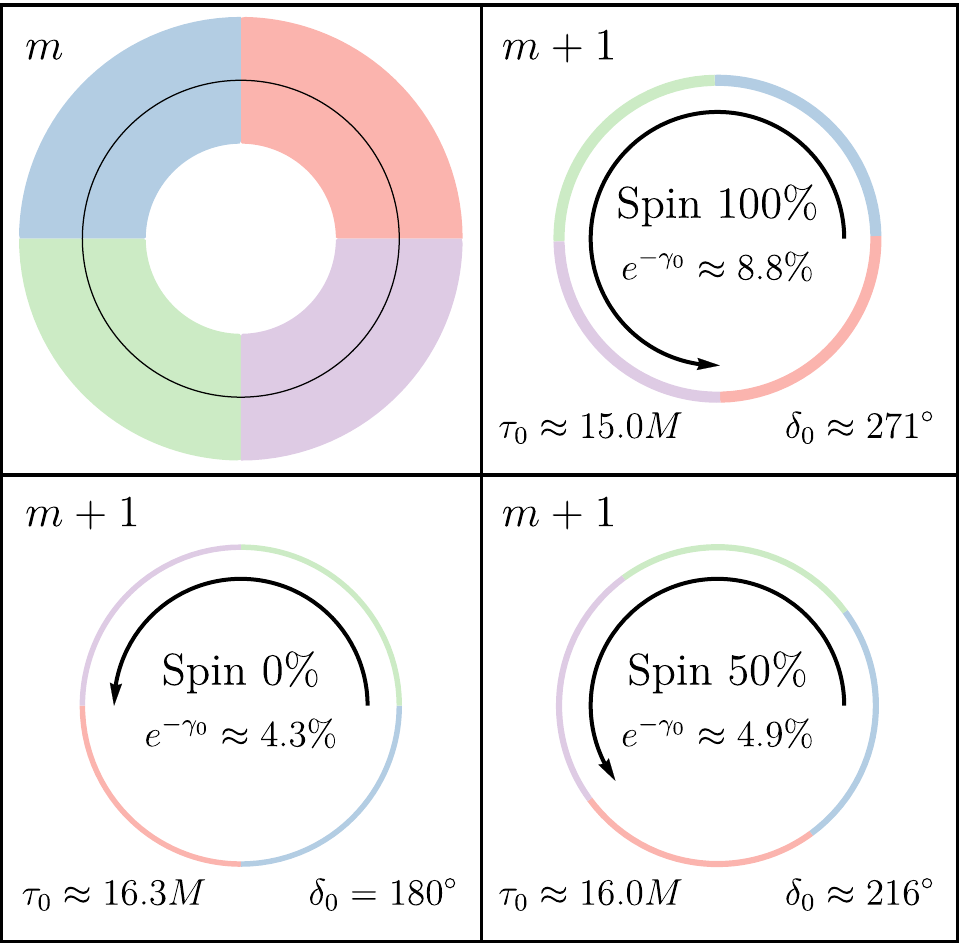}
	\caption{The critical parameters $\delta_0$, $\tau_0$, and $\gamma_0$ for an on-axis observer.  Above, we show their dependence on black hole spin, and below, we schematically illustrate their effects.  Successive images are demagnified by $e^{-\gamma_0}$, rotated by $\delta_0$, and delayed by $\tau_0$.  The image labeled $m$ (top left) is shown artificially large, but the demagnified images are then to scale.}
	\label{fig:CriticalExponents}
\end{figure}

\begin{figure*}
	\centering
	\includegraphics[width=\textwidth]{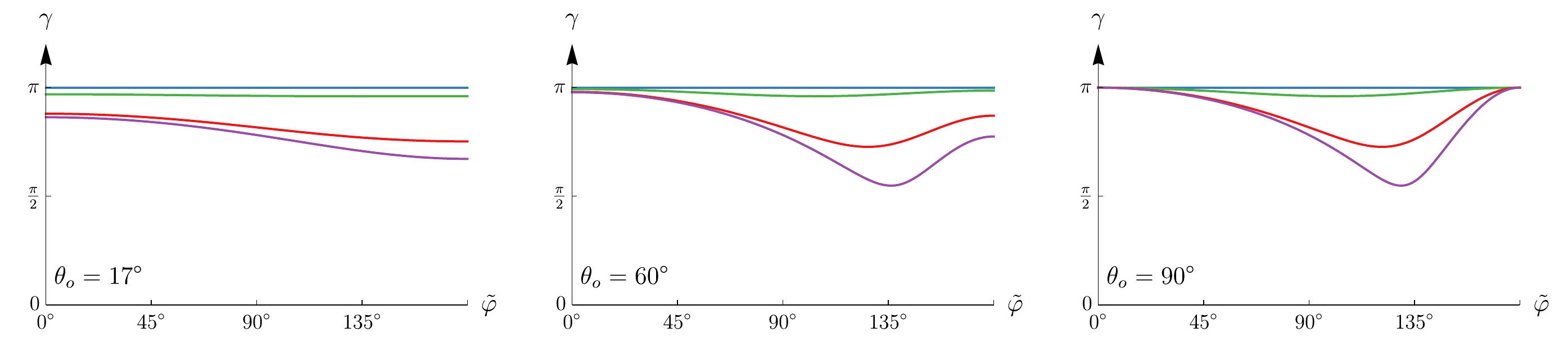}\\
	\includegraphics[width=\textwidth]{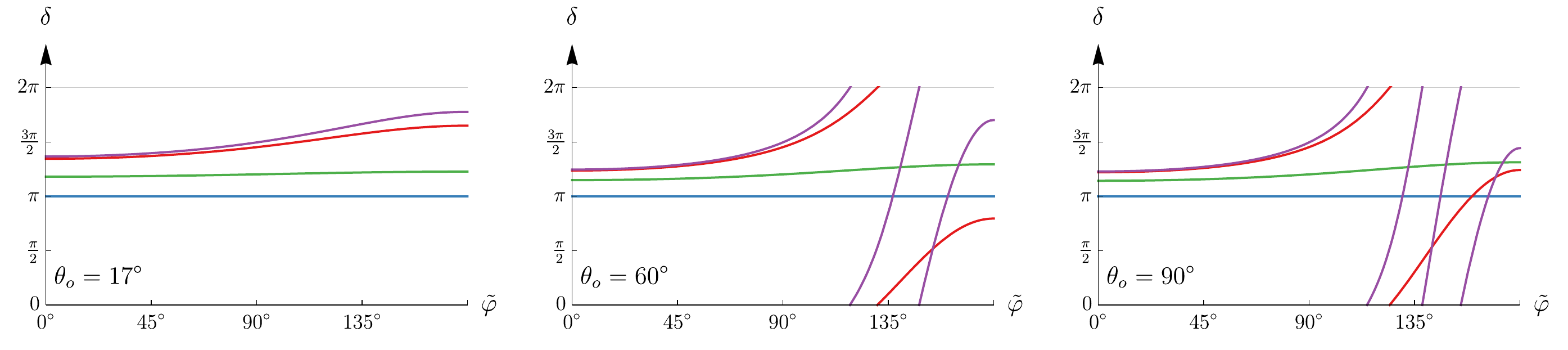}\\
	\includegraphics[width=\textwidth]{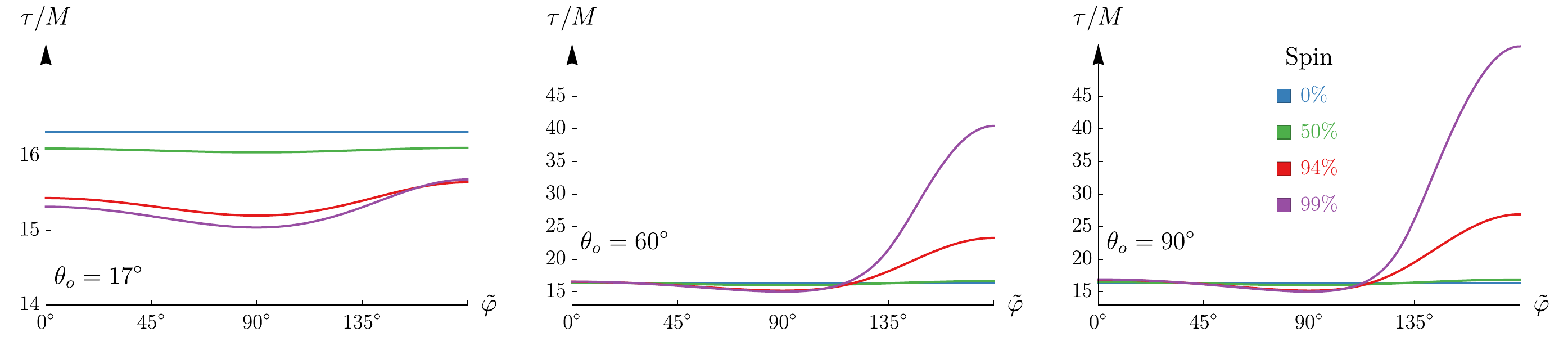}
	\caption{Variation of the critical parameters $\gamma$, $\delta$, and $\tau$ around the critical curve.  We show the value of these parameters as a function of polar angle $\tan{\tilde{\varphi}}=\tilde{\beta}/\tilde{\alpha}$ around the curve.  (For $\delta$, we plot modulo $2\pi$.)  The rotation and delay parameters $\delta$ and $\tau$ become large near the NHEKline (Fig.~\ref{fig:Screen}) present for inclined observers of rapidly rotating black holes.  The demagnification parameter $\gamma$ becomes small near the edges of the NHEKline (see also Fig.~6 of Ref.~\cite{Johnson2019}).  The time delay $\tau\sim16M$ has been seen previously in numerical simulations of emitting sources near black holes \cite{Fukumura2008,Moriyama2019}.}
	\label{fig:CriticalParameters}
\end{figure*}

\subsection{Equatorial sources viewed from the pole}

To unpack the physics of the photon ring, we begin with the simplest case of an equatorial source ($\theta_s=\pi/2$) and a polar observer ($\theta_o=0$).  In this case, the source integrals vanish ($f_s=\pi_s=e'_s=0$), and the observer integrals become complete ($f_o=\pi_o=e'_o=1$).  Together with $\pm_o=-1$ [Eq.~\eqref{eq:OnAxisSign}], this reduces Eq.~\eqref{eq:OrbitNumberTurningPoint} to
\begin{align}
	\label{eq:nEquatorialPolar}
	n\approx\frac{m}{2}+\frac{1}{4}.
\end{align}
Likewise, Eqs.~\eqref{eq:dLyapunov}, \eqref{eq:phiAsymptotic}, and \eqref{eq:tAsymptotic} simplify to
\begin{align}
	\label{eq:dSimple}
	d&\approx\frac{1}{C_\pm}\exp\br{-\pa{m+\frac{1}{2}}\gamma},\\
	\label{eq:phiSimple}
	\Delta\phi&\approx\pa{m+\frac{1}{2}}\hat{\delta}+D_\pm,\\
	\label{eq:tSimple}
	\Delta t&\approx\pa{m+\frac{1}{2}}\tau+H_\pm.
\end{align}
Because of the discontinuity in $\hat{\delta}$, at this stage, we consider $\theta_o$ to be small but finite.

The formulas \eqref{eq:dSimple}, \eqref{eq:phiSimple}, and \eqref{eq:tSimple} encode the arrival position and time of the infinitely many apparent positions of a given source.  The details are determined by the dependence of the coefficients $C_\pm$, $D_\pm$, and $H_\pm$ on the source radius $r_s$.  However, these terms are independent of the image number $m$, and hence cancel out of appropriate ratios and differences,\begin{align}
	\frac{d_{m+1}}{d_{m}}&\approx e^{-\gamma},\\
	\label{eq:phiPolarJump}
	(\Delta\phi)_{m+1}-(\Delta\phi)_m&\approx\hat{\delta},\\
	(\Delta t)_{m+1}-(\Delta t)_m&\approx\tau.  
\end{align}
We may now replace $\hat{\delta}$ with $\delta$ since the two agree modulo $2\pi$ [i.e., the difference can be absorbed into the left-hand side of Eq.~\eqref{eq:phiPolarJump}].  Then all quantities are continuous and we may take the full limit $\theta_o\to0$.  Recalling that $d=b-\tilde{b}$ and $\phi_o=\varphi$, and additionally denoting the observation time $t_o$ by $t$, we thus obtain
\begin{align}
	\label{eq:bPolar}
	\frac{b_{m+1}-\tilde{b}}{b_{m}-\tilde{b}}&\approx e^{-\gamma_0},\\
	\label{eq:phiPolar}
	\varphi_{m+1}-\varphi_m&\approx\delta_0,\\
	\label{eq:tPolar}
	t_{m+1}-t_m&\approx\tau_0, 
\end{align}
where $\gamma_0$, $\delta_0$, and $\tau_0$ were given in Eqs.~\eqref{eq:gamma0}, \eqref{eq:delta0}, and \eqref{eq:tau0}, respectively.

Eqs.~\eqref{eq:bPolar}, \eqref{eq:phiPolar} and \eqref{eq:tPolar} show that the successive apparent positions of a source $(r_s,\theta_s=\pi/2,\phi_s,t_s)$ move a factor of $e^{\gamma_0}$ closer to the critical curve for every additional half-orbit, while rotating an angle $\delta_0$ around the curve and appearing a time $\tau_0$ later.  Recalling that $\delta_0=\pi$ for a nonspinning black hole, we see that successive images appear on opposite sides of the critical curve.  This is easily understood from the geometry of the source (Fig.~\ref{fig:Backside} left). 

Note that the arrival positions can be neatly represented in terms of a complex coordinate $z=(b-\tilde{b})e^{i\varphi}$,\footnote{This coordinate maps the image plane to two copies of the complex plane, one inside $\mathcal{C}$ and one outside.} such that
\begin{align}
	z_{m+1}=e^{-\gamma_0+i\delta_0}z_m.
\end{align}
Thus we may view $-\gamma_0+i\delta_0$ as a single complex exponent.

Now consider an equatorial source of some finite extent from $r^{-}_{s}$ to $r^{+}_{s}$, as in the emitting portion of an accretion disk.  Let $b_m^\pm(\tilde{r})$ represent the $m^\text{th}$ observed position of the inner and outer edges.  At some sufficiently high $m$ (typically $m\geq1$ is sufficient), we may compute $b_m^\pm$ using the approximation \eqref{eq:bPolar}.  Denoting the apparent width of each image by $\Delta b_m =b_m^{+}-b_m^{-}$, from Eq.~\eqref{eq:dSimple} we have
\begin{align}
	\frac{\Delta b_{m+1}}{\Delta b_m}\approx e^{-\gamma_0}.
\end{align}
That is, successive images of the equatorial disk are \textit{demagnified} (narrower) by a factor of $e^{-\gamma_0}$.  The total flux associated with each image also decreases by the same typical factor, i.e., the flux is exponentially suppressed in the orbit number.  Each successive image also rotates on the screen by an angle $\delta_0$, an effect which would be visible for nonaxisymmetric source profiles.  Finally, each successive image arrives a time $\tau_0$ later, an effect that would be observable for time-variable source profiles.  Some of these properties are illustrated in Fig.~\ref{fig:CriticalExponents}.

\subsection{General sources viewed from the pole}
\label{sec:GeneralSourcePolarObserver}

Suppose now that the source is not equatorial, but the observer is still on the pole.  From Eq.~\eqref{eq:OrbitNumberTurningPoint} using $\pm_o=-1$ and $f_o=\pi_o=e'_o=1$ yet again, we have
\begin{align}
	n=\frac{m}{2}+\frac{1}{4}-\frac{(-1)^m}{4}f_s.
\end{align}
Repeating the same procedure that led to Eqs.~\eqref{eq:bPolar}, \eqref{eq:phiPolar} and \eqref{eq:tPolar}, we now find\footnote{As before, one may take the $\theta_o\to0$ limit only after having eliminated all dependence on $\sign(\tilde{\lambda})=-\sign(\tilde{\alpha})=-\sign(\tilde{r}-\tilde{r}_0)$.  This direction-dependence in the limit enters via the discontinuous quantity $\hat{\delta}$ as well as the discontinuous limit \eqref{eq:PoleJump}.  Note also that $\pi_s$ vanishes in the limit from both sides.}
\begin{align}
	\label{eq:bPolarGeneral}
	\frac{b_{m+1}-\tilde{b}}{b_m-\tilde{b}}&\approx e^{-x_m\gamma_0},\\
	\label{eq:phiPolarGeneral}
	\varphi_{m+1}-\varphi_m&\approx x_m\delta_0-(-1)^m\pi f_s.\\
	\label{eq:tPolarGeneral}
	t_{m+1}-t_m&\approx x_m\tau_0+(-1)^m\frac{4a\tilde{u}_+\tilde{E}'}{\sqrt{-\tilde{u}_-}}\pa{f_s-e'_s},
\end{align}
with
\begin{align}
	x_m=1+(-1)^mf_s.
\end{align}
Thus, although $\gamma_0$, $\delta_0$, and $\tau_0$ no longer give precisely the demagnification, rotation, and time delay (respectively), they still encode these effects in a relatively straightforward way, depending on whether $m$ is even or odd.  We again obtain simple expressions if we advance $m$ by two instead of one,
\begin{align}
	\frac{b_{m+2}-\tilde{b}}{b_m-\tilde{b}}&\approx e^{-2\gamma_0},\\
	\varphi_{m+2}-\varphi_m&\approx2\delta_0,\\
	t_{m+2}-t_m&\approx2\tau_0.
\end{align}
Thus, a given source gives rise to two families of images (one for even $m$ and one for odd $m$), each of which has demagnification $2\gamma_0$, rotation $2\delta_0$, and time delay $2 \tau_0$.  These are just the Lyapunov exponent, lapse in $\phi$, and lapse in $t$ for a complete bound photon orbit, respectively.  That is, each successive image of each family differs by one orbit around the black hole.  Roughly speaking, the two families correspond to emission towards and away from the observer; for an equatorial disk, they are images of the front and back of the disk, respectively.

Recall that $\delta_0=\pi$ in Schwarzschild.  As such, each family of images approaches the critical curve radially, since the rotation of each successive image is $2\pi\sim0$.

\subsection{Inclined observer: Equatorial sources}

Next, suppose that the source is equatorial ($\theta_s=\pi/2$), so that $f_s=0$, while the observer is inclined ($\theta_o\neq0$), so that $\pm_o=\sign\pa{\beta}$.  Then Eq.~\eqref{eq:OrbitNumberTurningPoint} becomes
\begin{align}
	n=\frac{m}{2}-\frac{1}{4}\sign\pa{\beta}f_o.
\end{align}
Recalling that we set $\phi_o=0$ for the inclined observer, it follows from Eqs.~\eqref{eq:dLyapunov}, \eqref{eq:phiAsymptotic}, and \eqref{eq:tAsymptotic} that
\begin{align}
	\label{eq:dEquatorial}
	d&\approx\frac{1}{C_\pm}\exp\br{-\gamma\pa{m-\frac{1}{2}\sign\pa{\beta}f_o}},\\
	\phi_s&\approx-\pa{m-\frac{1}{2}\sign\pa{\beta}f_o}\hat{\delta}\nonumber\\
	\label{eq:phiEquatorial}
	&\quad-\sign\pa{\beta}\frac{\tilde{\lambda}\tilde{\Pi}}{a\sqrt{-\tilde{u}_-}}\pa{f_o-\pi_o}-D_\pm,\\
	t-t_s&\approx\pa{m-\frac{1}{2}\sign\pa{\beta}f_o}\tau\nonumber\\
	\label{eq:tEquatorial}
	&\quad-\sign\pa{\beta}\frac{(2 a \tilde{u}_+ \tilde{E}')}{\sqrt{-\tilde{u}_-}}(f_o-e'_o) + H_\pm.
\end{align}
As our observer is now inclined, the quantities $\gamma$, $\hat{\delta}$, and $\tau$ depend nontrivially on $\tilde{r}$, which together with the sign of $\beta$ specifies a point on the critical curve.  Selecting a position $(\tilde{r},\sign\pa{\beta})$ on the critical curve, Eq.~\eqref{eq:dEquatorial} gives the perpendicular distance of a photon that originated at $(r_s,\theta_s=\pi/2)$ and encountered $m$ polar turning points on its way.  The emission angle $\phi_s$ of this photon is given by Eq.~\eqref{eq:phiEquatorial}, and the emission time $t_s$ by Eq.~\eqref{eq:tEquatorial} (in terms of the observation time $t_o=t$). 

We may again take a ratio to find
\begin{align}
	\label{eq:dm-eq}
	\frac{d_{m+1}}{d_m}\approx e^{-\gamma},
\end{align}
which may be compared with \eqref{eq:bPolar} above.  Fixing a position $(\tilde{r},\sign \beta)$ along the critical curve, Eq.~\eqref{eq:dm-eq} shows that photons from a given equatorial source ring $(r_s,\theta_s=\pi/2)$ arrive at perpendicular distances $d$ that successively decrease by a factor of $e^{-\gamma}$.  Fixing the observation time $t$, these photons originated from angles $\phi_s^m$ and times $t_s^m$ related by
\begin{align}
	\label{eq:phim-eq}
	\phi_s^{m+1}-\phi_s^m&\approx-\delta,\\
	\label{eq:tm-eq}
	t_s^{m+1}-t_s^m&\approx-\tau,
\end{align}
where now we have switched to the continuous quantity $\delta$, absorbing the jump of $2\pi$ into the $\phi$ coordinate.

Recall that $\gamma$, $\delta$, and $\tau$ depend on the critical curve position $\tilde{r}$ under consideration.  For a stationary, axisymmetric source, we may regard $e^{-\gamma}$ as a demagnification factor that varies over the critical curve.  For a general equatorial source, we see no simple way to describe the properties of the images in terms of those of the source, but it is clear from the exceptionally simple formulas \eqref{eq:dm-eq}, \eqref{eq:phim-eq}, and \eqref{eq:tm-eq} that $\gamma$, $\delta$, and $\tau$ still encode universal features of high-order images.  The variation of these critical parameters is shown in Fig.~\ref{fig:CriticalParameters}.

\subsection{General source and observer}

For nonequatorial sources sources observed at nonzero inclination, Eqs.~\eqref{eq:dEquatorial}, \eqref{eq:phiEquatorial} and \eqref{eq:tEquatorial} are supplemented by terms involving dependence on $m$ through $(-1)^m$, as in Eqs.~\eqref{eq:bPolarGeneral}, \eqref{eq:phiPolarGeneral}, and \eqref{eq:tPolarGeneral} above.  These terms give rise to separate behavior for even and odd values of $m$, as described in Sec.~\ref{sec:GeneralSourcePolarObserver} above in the case of a polar observer.  Rather than present these details, we instead merely note that in the general case we still have simple expressions when $m$ is shifted by two,
\begin{align}
	\label{eq:dm}
	\frac{d_{m+2}}{d_m}&\approx e^{-2\gamma},\\
	\label{eq:phim}
	\phi_s^{m+2}-\phi_s^{m}&\approx-2\delta,\\
	\label{eq:tm}
	t_s^{m+2}-t_s^m&\approx-2\tau.
\end{align}
That is, given any source ring $(r_s,\theta_s)$ observed at any inclination $\theta_o$ at some time $t$, and choosing any perpendicular $(\tilde{r},\sign\pa{\beta})$ to the image-plane critical curve, photons arrive in two separate families (even and odd $m$) at distances decreasing by factors of $e^{-2\gamma}$, which were emitted at successively earlier times (with delay $-2\tau$) as well as different positions around the ring (with increment $-2\delta$).  Although these properties do not translate in any simple way into a description of the distortion and demagnification of a general source observed at a general inclination, it is clear from the exceptionally simple formulas \eqref{eq:dm}, \eqref{eq:phim}, and \eqref{eq:tm} that $\gamma$, $\delta$, and $\tau$ still encode universal features of high-order images.

\acknowledgements{SEG was supported in part by NSF grant PHY-1752809 to the University of Arizona.  Portions of this work were completed at the Aspen Center for Physics, which is supported by National Science Foundation grant PHY-1607611.  AL was supported in part by the Jacob Goldfield Foundation.}

\appendix

\section{Radial roots and integrals}
\label{app:RadialAnalysis}

In Ref.~\cite{KerrGeodesics}, we derived analytic formulas for the roots of the radial potential \eqref{eq:RadialPotential} which are ordered when the roots are real.  We reproduce these formulas here for convenience.  We introduce 
\begin{align}
	\mathcal{A}&=a^2-\eta-\lambda^2,\\
	\mathcal{B}&=2M\br{\eta+\pa{\lambda-a}^2} >0,\\
	\mathcal{C}&=-a^2\eta,
\end{align}
and further define 
\begin{align}
	\mathcal{P}&=-\frac{\mathcal{A}^2}{12}-\mathcal{C},\\
	\mathcal{Q}&=-\frac{\mathcal{A}}{3}\br{\pa{\frac{\mathcal{A}}{6}}^2-\mathcal{C}}-\frac{\mathcal{B}^2}{8},
\end{align} 
as well as
\begin{align}
	z&=\sqrt{\frac{\omega_++\omega_-}{2}-\frac{\mathcal{A}}{6}}
	>0,\\
	\omega_\pm&=\sqrt[3]{-\frac{\mathcal{Q}}{2}\pm\sqrt{\pa{\frac{\mathcal{P}}{3}}^3+\pa{\frac{\mathcal{Q}}{2}}^2}}.
\end{align}
The four roots are then given by
\begin{subequations}
\label{eq:RadialRoots}
\begin{align}
	r_1&=-z-\sqrt{-\frac{\mathcal{A}}{2}-z^2+\frac{\mathcal{B}}{4z}}, \\
	r_2&=-z+\sqrt{-\frac{\mathcal{A}}{2}-z^2+\frac{\mathcal{B}}{4z}},\\
	r_3&=z-\sqrt{-\frac{\mathcal{A}}{2}-z^2-\frac{\mathcal{B}}{4z}},\\
	\label{eq:r4}
	r_4&=z+\sqrt{-\frac{\mathcal{A}}{2}-z^2-\frac{\mathcal{B}}{4z}}.
\end{align}
\end{subequations}
These roots always satisfy $r_i \leq r_j$ when $i<j$ and both $r_i$ and $r_j$ are real.  On the critical curve $\mathcal{C}$, we have $r_3=r_4$, but otherwise $r_4$ is always the largest real root outside the horizon.  Thus, rays reaching infinity either have a turning point at $r_4$, are asymptotic to a photon orbit at $r_3=r_4$, or have no turning point at all (when $r_4$ is complex, or real but inside the horizon).

We now present the results from Ref.~\cite{KerrGeodesics} needed to compute the radial integrals of interest to this paper.  Rays that arrive outside the critical curve are case (2) of Ref.~\cite{KerrGeodesics}.  The antiderivative is given by Eqs.~(B35)-(B40) therein,
\begin{align}
	\label{eq:Case2Antiderivative}
	\mathcal{I}_r^{(2)}(r)=\frac{2}{\sqrt{r_{31}r_{42}}}F\pa{\left.\arcsin\sqrt{\frac{r-r_4}{r-r_3}\frac{r_{31}}{r_{41}}}\right|\frac{r_{32}r_{41}}{r_{31}r_{42}}}.
\end{align}
In particular, the complete radial integral \eqref{eq:IrTotal} is
\begin{align}
	I_r^\mathrm{total}=\frac{4}{\sqrt{r_{31}r_{42}}}F\pa{\left.\arcsin\sqrt{\frac{r_{31}}{r_{41}}}\right|\frac{r_{32}r_{41}}{r_{31}r_{42}}}.
\end{align}
Rays that arrive inside the critical curve are also case (2) when all roots are real; otherwise, if $r_3=\bar{r}_4$ are complex conjugate roots, then the rays are case (3).  For case (2), the antiderivative is again Eq.~\eqref{eq:Case2Antiderivative}, whereas for case (3), the antiderivative is given by Eqs.~(B55) and (B67)-(B71) of Ref.~\cite{KerrGeodesics},
\begin{align}
	\mathcal{I}_r^{(3)}(r)&=\frac{1}{\sqrt{AB}}F\pa{\left.\arccos{\frac{A\pa{r-r_1}-B\pa{r-r_2}}{A\pa{r-r_1}+B\pa{r-r_2}}}\right|k_3},\nonumber\\
	A&=\sqrt{r_{32}r_{42}}
	>0,\quad
	B=\sqrt{r_{31}r_{41}}
	>0,\\
	k_3&=\frac{\pa{A+B}^2-r_{21}^2}{4AB}
	\in(0,1).
\end{align}
In particular, the complete radial integral \eqref{eq:IrTotal} is
\begin{align}
	I_r^\mathrm{total}=\frac{2}{\sqrt{r_{31}r_{42}}}F\pa{\left.\arcsin\sqrt{\frac{r_{31}}{r_{41}}}\right|\frac{r_{32}r_{41}}{r_{31}r_{42}}}-\mathcal{I}_r^{(2)}(r_+),
\end{align}
if all roots are real; otherwise, when $r_3=\bar{r}_4$, it is 
\begin{align}
	I_r^\mathrm{total}&=\frac{1}{\sqrt{AB}}F\pa{\left.\arccos{\frac{A-B}{A+B}}\right|k_3}-\mathcal{I}_r^{(3)}(r_+).
\end{align}

\onecolumngrid

\section{Asymptotic approximation for the radial integrals}
\label{app:MAE}

The integrands of the fundamental radial integrals $I_r$, $I_\phi$, and $I_t$ involve (the square root of) the radial potential \eqref{eq:RadialPotential} in their denominators.  Single roots of $\mathcal{R}(r)$ correspond to turning points where the integral remains finite.  However, for critical conserved quantities $\lambda=\tilde{\lambda}(\tilde{r})$ and $\eta=\tilde{\eta}(\tilde{r})$, the roots $r_3$ and $r_4$ [Eqs.~\eqref{eq:RadialRoots}] coalesce, rendering the integral logarithmically divergent at the double root $\tilde{r}=r_3=r_4$.  Physically, this represents a critical photon asymptotically approaching its associated photon orbit radius $\tilde{r}$.  If the conserved quantities are not precisely critical but only nearly so, then the total integral $I_r$ is finite for each such ray, but the value diverges logarithmically in the deviation of the conserved quantities from their critical values.  Physically, this represents a near-critical photon spending an asymptotically large amount of time orbiting near its associated bound photon orbit at $\tilde{r}$.  In this situation, one expects the integral to break into two contributions, one from near the photon orbit and one from far away, such that the near-critical integral can be estimated by the method of matched asymptotic expansions.  In this appendix, we compute the relevant approximations to the radial integrals using this method.

All bound photon orbits cross the equatorial plane and hence have $\eta>0$.  Therefore, in this appendix, we will use
\begin{align}
	q=\sqrt{\eta}
	>0,
\end{align}
in lieu of $\eta$.  Consider a null geodesic whose conserved quantities $\lambda$ and $q$ are nearly equal to those of a bound photon orbit.  We may then write
\begin{align}
	\label{eq:Variation}
	\lambda=\tilde{\lambda}\pa{1+\dt\lambda},\qquad
	q=\tilde{q}\pa{1+\dt q},
\end{align}
where $\tilde{\lambda}$ and $\tilde{\eta}=\tilde{q}^2$ are associated to the radius $\tilde{r}$ of the photon orbit by  Eqs.~\eqref{eq:lambdaCritical} and \eqref{eq:etaCritical}, and $\ab{\dt\lambda}\sim\ab{\dt q}\ll1$.  We also introduce a new radial coordinate $\dt r$ by
\begin{align}
	r=\tilde{r}\pa{1+\dt r},
\end{align}
and use it to define ``near'' and ``far'' zones as follows:
\begin{align}
	&\text{Near:}&&\ab{\dt r}\ll1,\\
	&\text{Far:}&&\ab{\dt\lambda}\sim\ab{\dt q}\ll\ab{\dt r}.
\end{align}
These regimes overlap when $\ab{\dt\lambda}\sim\ab{\dt q}\ll\ab{\dt r}\ll1$.  We emphasize that throughout this discussion, ``near'' and ``far'' refer to distance from the photon orbit radius $\tilde{r}$, rather than distance from the black hole.  The far-zone region is disjoint, consisting of a ``right'' region containing asymptotic infinity, and a ``left'' region containing the event horizon.

The radial potential $\mathcal{R}(r)$ has different approximations in the near and far zones.  In the near-zone, it is well approximated by the scaling regime $\dt r^2\sim\dt\lambda\sim\dt q$, in which
\begin{align}
	\mathcal{R}(r)\approx\mathcal{R}_n(\dt r)
	:=4\tilde{r}^4\tilde{\chi}\pa{\dt r^2-\dt r_0^2},
\end{align}
where $\tilde{\chi}$ is as defined in Eq.~\eqref{eq:chi}, and in the last step we also introduced a dimensionless quantity\footnote{A related quantity $\dt B=\tilde{\chi}\dt r_0^2$ was introduced previously in Ref.~\cite{Gralla2019}.  Here, we present it in greatly simplified form and, importantly, show that it is proportional to the perpendicular distance $\ab{d}$ from the curve $\mathcal{C}$ [see Eq.~\eqref{eq:Distance}].}
\begin{align}
	\label{eq:Fr}
	\dt r_0^2&=\frac{\Delta(\tilde{r})}{2\tilde{r}^2\tilde{\chi}}\br{-\pa{\frac{\tilde{r}-3M}{\tilde{r}-M}}\frac{\tilde{\lambda}}{a}\dt\lambda+\frac{\tilde{q}^2}{\tilde{r}^2}\dt q},\qquad
	\tilde{\chi}=1-\frac{M\Delta(\tilde{r})}{\tilde{r}\pa{\tilde{r}-M}^2}.
\end{align}
Notice that the quadratic near-zone potential $\mathcal{R}_n(\dt r)$ has zeros at $\dt r=\pm\dt r_0$; these correspond to radial turning points provided that $\dt r_0^2>0$.  For photons that reach infinity, only the outer root is relevant.  Note also that
\begin{align}
	\tilde{\chi}=\frac{3}{4}-\pa{\frac{a\tilde{q}}{2\tilde{r}^2}}^2\in\left(0,\frac{3}{4}\right].
\end{align}

In the far-zone, the radial potential $\mathcal{R}(r)$ is instead well-approximated by its value at $\lambda=\tilde{\lambda}$ and $q=\tilde{q}$,
\begin{align}
	\mathcal{R}(r)\approx\mathcal{R}_f(\dt r)
	:=4\tilde{r}^4\dt r^2\pa{\frac{\dt r^2}{4}+\dt r+\tilde{\chi}}
	=4\tilde{r}^4\tilde{\chi}\dt r^2\mathcal{Q}(\dt r),
\end{align}
where in the last step, we introduced for future convenience a function
\begin{align}
	\mathcal{Q}(\dt r)=1+\frac{\dt r}{\tilde{\chi}}+\frac{\dt r^2}{4\tilde{\chi}}.
\end{align}
The double root $\dt r=0$ of $\mathcal{R}_f(\dt r)$ is outside the regime of validity of the far-zone approximation and does not correspond to a physical turning point.  (It is the far-zone remnant of the two roots $\dt r=\pm\dt r_0$ that are separately resolved by the near-zone approximation.)  The quartic potential $\mathcal{R}_f(\dt r)$ has two other negative roots $\dt r_0^-<\dt r_0^+<0$, where $\dt r_0^\pm=2\pa{-1\pm\sqrt{1-\tilde{\chi}}}$, which a photon that comes in from infinity cannot encounter.

If a light ray with conserved quantities \eqref{eq:Variation} reaches infinity, then by definition it arrives near the closed curve $\mathcal{C}$.  Rays arriving inside $\mathcal{C}$ have no radial turning points, while rays arriving outside have a single turning point.  The preceding analysis shows that these cases correspond to $\dt r^2_0<0$ and $\dt r_0^2>0$, respectively:
\begin{align}
	&\text{Inside $\mathcal{C}$:}&&
	\dt r_0^2<0,\\
	&\text{Outside $\mathcal{C}$:}&&
	\dt r_0^2>0.
\end{align}
In Sec.~\ref{app:Distance} below, we show that $\dt r_0^2$ is actually proportional to the (signed) perpendicular distance from $\mathcal{C}$.

We have now laid the groundwork to compute the geodesic path integrals involving the radial potential. To do so, it will suffice to evaluate the definite integrals
\begin{align}
	\label{eq:RadialIntegrals}
	I_r^{ab}=\int_{r_a}^{r_b}\frac{\ed r}{\sqrt{\mathcal{R}(r)}},\qquad
	I_\phi^{ab}=\int_{r_a}^{r_b}\frac{a\pa{2Mr-a\lambda}}{\Delta(r)\sqrt{\mathcal{R}(r)}}\ed r,\qquad
	I_t^{ab}=\int_{r_a}^{r_b}\frac{r^2\Delta(r)+2Mr\pa{r^2+a^2-a\lambda}}{\Delta(r)\sqrt{\mathcal{R}(r)}}\ed r,
\end{align}
for all combinations of in/out for the conserved quantities, and near/far for each of $r_a$ and $r_b$.  We will organize the calculation in sections based on the near/far split, considering only the cases that arise when photons reach infinity.  We will present integrals in terms of the inverse hyperbolic tangent, defined as
\begin{align}
	\arctanh{x}=\frac{1}{2}\log\pa{\frac{1+x}{1-x}},
\end{align}
which is manifestly real whenever $x\in\br{-1,1}$.

\subsection{Both points in the near-zone}
\label{sec:NearNear}

When both endpoints of the geodesic are in the near-zone, the radial integrals \eqref{eq:RadialIntegrals} simplify to
\begin{align}
	I_r^\mathrm{nn}=\frac{1}{2\tilde{r}\sqrt{\tilde{\chi}}}\int_{\dt r_a}^{\dt r_b}\frac{\ed(\dt r)}{\sqrt{\dt r^2-\dt r_0^2}},\qquad
	I_\phi^\mathrm{nn}=a\pa{\frac{\tilde{r}+M}{\tilde{r}-M}}I_r^\mathrm{nn},\qquad
	I_t^\mathrm{nn}=\tilde{r}^2\pa{\frac{\tilde{r}+3M}{\tilde{r}-M}}I_r^\mathrm{nn},
\end{align}
where the label ``nn'' stands for ``near-near'' (i.e., both points in the near-zone).  Here and hereafter, $\dt r_a$ and $\dt r_b$ denote the $\dt r$-coordinate values of the Boyer-Lindquist radii $r_a$ and $r_b$, respectively, with $\dt r_b<\dt r_b$.

We can now evaluate $I_r^\mathrm{nn}$.  Photons arriving outside $\mathcal{C}$ have $0<\dt r_0\le\dt r$ and the manifestly real integral
\begin{align}
	\label{eq:NearIntegralOut}
	I_r^\mathrm{nn,out}(\dt r_a,\dt r_b)=\frac{1}{2\tilde{r}\sqrt{\tilde{\chi}}}\left.\arctanh\pa{\frac{\sqrt{\dt r^2-\dt r_0^2}}{\dt r}}\right|_{\dt r_a}^{\dt r_b}.
\end{align}
On the other hand, photons arriving inside $\mathcal{C}$ have $\dt r_0^2<0$ and the manifestly real integral
\begin{align}
	I_r^\mathrm{nn,in}(\dt r_a,\dt r_b)=\frac{1}{2\tilde{r}\sqrt{\tilde{\chi}}}\left.\arctanh\pa{\frac{\dt r}{\sqrt{\dt r^2-\dt r_0^2}}}\right|_{\dt r_a}^{\dt r_b}.
\end{align}
These results can be combined into a single formula
\begin{align}
	I_r^\mathrm{nn}(\dt r_a,\dt r_b)=\left.\frac{\sign\pa{\dt r}}{4\tilde{r}\sqrt{\tilde{\chi}}}\log\pa{\sign\pa{\dt r_0^2}\frac{1+\sqrt{1-\frac{\dt r_0^2}{\dt r^2}}}{1-\sqrt{1-\frac{\dt r_0^2}{\dt r^2}}}}\right|_{\dt r_a}^{\dt r_b}.
\end{align}

\subsection{Both points in one region of the far-zone}
\label{sec:FarFar}

The far-zone consists of two disjoint regions on either side of the near-zone, with one region containing the horizon, and the other region containing asymptotic infinity.  When both points endpoints of the geodesic are in the same region of the far-zone, the radial integrals \eqref{eq:RadialIntegrals} reduce to
\begin{align}
	I_r^\mathrm{ff}&=\frac{1}{2\tilde{r}\sqrt{\tilde{\chi}}}\int_{\dt r_a}^{\dt r_b}\frac{\ed(\dt r)}{\sqrt{\dt r^2\mathcal{Q}(\dt r)}},\\
	I_\phi^\mathrm{ff}&=\frac{aM}{\tilde{r}^2\sqrt{\tilde{\chi}}}\int_{\dt r_a}^{\dt r_b}\frac{c_0+\pa{1+\dt r}}{\pa{\dt r-\dt r_+}\pa{\dt r-\dt r_-}}\frac{\ed(\dt r)}{\sqrt{\dt r^2\mathcal{Q}(\dt r)}},\\
	I_t^\mathrm{ff}&=\frac{\tilde{r}}{2\sqrt{\tilde{\chi}}}\int_{\dt r_a}^{\dt r_b}\frac{c_1\pa{1+\dt r}+c_2\pa{1+\dt r}^2+\pa{1+\dt r}^4}{\pa{\dt r-\dt r_+}\pa{\dt r-\dt r_-}}\frac{\ed(\dt r)}{\sqrt{\dt r^2\mathcal{Q}(\dt r)}},
\end{align}
where we introduced dimensionless coefficients
\begin{align}
	c_0=-\frac{a\tilde{\lambda}}{2M\tilde{r}},\qquad
	c_1=\frac{2aM}{\tilde{r}^3}\pa{a-\tilde{\lambda}},\qquad
	c_2=\frac{a^2}{\tilde{r}^2},
\end{align}
and $\dt r_\pm$ denotes the $\dt r$-coordinate of the outer/inner event horizon,
\begin{align}
	\label{eq:dtrpm}
	\dt r_\pm&=\frac{M\pm\sqrt{M^2-a^2}}{\tilde{r}}-1\in\pa{-1,0}.
\end{align}
Now define a symmetric function of two variables
\begin{align}
	\label{eq:Q2}
	\mathcal{Q}_2(\dt r_a,\dt r_b)=\frac{2\sqrt{\mathcal{Q}(\dt r_a)}\sqrt{\mathcal{Q}(\dt r_b)}}{\mathcal{Q}(\dt r_a)+\mathcal{Q}(\dt r_b)-\frac{\pa{\dt r_a-\dt r_b}^2}{4\tilde{\chi}}}
	\in(0,1],
\end{align}
whose range $(0,1]$, which assumes that both $\dt r_a$ and $\dt r_b$ are outside the event horizon $\dt r_+$ (but not that they are positive), is derived in Sec.~\ref{sec:Range} below. This range guarantees that the following functions are manifestly real outside the horizon:
\begin{align}
	\mathcal{Q}_\phi(\dt r)&=\frac{c_0+\pa{1+\dt r_+}}{\dt r_+\pa{\dt r_+-\dt r_-}\sqrt{\mathcal{Q}(\dt r_+)}}\arctanh{\mathcal{Q}_2(\dt r,r_+)}-\frac{c_0+\pa{1+\dt r_-}}{\dt r_-\pa{\dt r_+-\dt r_-}\sqrt{\mathcal{Q}(\dt r_-)}}\arctanh{\mathcal{Q}_2(\dt r,r_-)},\\
	\mathcal{Q}_t(\dt r)&=-4\tilde{\chi}\sqrt{\mathcal{Q}(\dt r)}-\frac{4M\sqrt{\tilde{\chi}}}{\tilde{r}}\arctanh{\mathcal{Q}_2(\dt r,\infty)}+\frac{c_1\pa{1+\dt r_+}+c_2\pa{1+\dt r_+}^2+\pa{1+\dt r_+}^4}{\dt r_+\pa{\dt r_+-\dt r_-}\sqrt{\mathcal{Q}(\dt r_+)}}\arctanh{\mathcal{Q}_2(\dt r,r_+)}\nonumber\\
	&\qquad-\frac{c_1\pa{1+\dt r_-}+c_2\pa{1+\dt r_-}^2+\pa{1+\dt r_-}^4}{\dt r_-\pa{\dt r_+-\dt r_-}\sqrt{\mathcal{Q}(\dt r_-)}}\arctanh{\mathcal{Q}_2(\dt r,r_-)}.
\end{align}
Manifestly real forms of the far integrals are then
\begin{align}
	I_r^\mathrm{ff}(\dt r_a,\dt r_b)&=\left.-\frac{\sign\pa{\dt r}}{2\tilde{r}\sqrt{\tilde{\chi}}}\arctanh{\mathcal{Q}_2(\dt r,0)}\right|_{\dt r_a}^{\dt r_b}\\
	&=\left.-\frac{\sign\pa{\dt r}}{2\tilde{r}\sqrt{\tilde{\chi}}}\arctanh\pa{\frac{\sqrt{\mathcal{Q}(\dt r)}}{1+\frac{\dt r}{2\tilde{\chi}}}}\right|_{\dt r_a}^{\dt r_b},\nonumber\\
	I_\phi^\mathrm{ff}(\dt r_a,\dt r_b)&=\left.-\frac{\sign\pa{\dt r}aM}{\tilde{r}^2\sqrt{\tilde{\chi}}}\br{\frac{\tilde{r}}{2M}\pa{\frac{\tilde{r}+M}{\tilde{r}-M}}\arctanh{\mathcal{Q}_2(\dt r,0)}+\mathcal{Q}_\phi(\dt r)}\right|_{\dt r_a}^{\dt r_b}\\
	&=a\pa{\frac{\tilde{r}+M}{\tilde{r}-M}}I_r^\mathrm{ff}(\dt r_a,\dt r_b)-\left.\frac{\sign\pa{\dt r}aM}{\tilde{r}^2\sqrt{\tilde{\chi}}}\mathcal{Q}_\phi(\dt r)\right|_{\dt r_a}^{\dt r_b},\nonumber\\
	I_t^\mathrm{ff}(\dt r_a,\dt r_b)&=-\left.\frac{\sign\pa{\dt r}\tilde{r}}{2\sqrt{\tilde{\chi}}}\br{\frac{\tilde{r}+3M}{\tilde{r}-M}\arctanh{\mathcal{Q}_2(\dt r,0)}+\mathcal{Q}_t(\dt r)}\right|_{\dt r_a}^{\dt r_b}\\
	&=\tilde{r}^2\pa{\frac{\tilde{r}+3M}{\tilde{r}-M}}I_r^\mathrm{ff}(\dt r_a,\dt r_b)-\left.\frac{\sign\pa{\dt r}\tilde{r}}{2\sqrt{\tilde{\chi}}}\mathcal{Q}_t(\dt r)\right|_{\dt r_a}^{\dt r_b}.\nonumber
\end{align}

\subsection{One point in the near-zone and one point in the far-zone}

We now wish to consider the case where one point is in the near-zone and the other point is in the far-zone.  This requires the method of matched asymptotic expansions, which we implement as follows.  First, we choose an arbitrary matching radius $\dt R$.  We then split the integral into a portion from $\dt r_a$ to $\dt R$, and a remaining portion from $\dt R$ to $\dt r_b$.  The arbitrary point $\dt R$ is assumed to be in the overlap region $\ab{\dt\lambda}\sim\ab{\dt q}\ll\ab{\dt R}\ll 1$, so that the first integral may be computed with the near-zone approximation (presented in Sec.~\ref{sec:NearNear}), while the second integral may be computed with the far-zone approximation (presented in Sec.~\ref{sec:FarFar}).  Using the relevant definite integrals computed in these sections, and taking into account their various approximations, the arbitrary radius $\dt R$ disappears from the final expressions.

We begin with $I_r$. Photons arriving outside $\mathcal{C}$ necessarily have $0<\dt r_0<\dt r_a\ll 1$ and $\dt r_a\ll\dt r_b$, and the answer is 
\begin{align}
	\label{eq:OutsideMatching}
	I_r^\mathrm{nf,out}(\dt r_a,\dt r_b)=-\frac{1}{2\tilde{r}\sqrt{\tilde{\chi}}}\br{\arctanh\pa{\frac{\sqrt{\mathcal{Q}(\dt r_b)}}{1+\frac{\dt r_b}{2\tilde{\chi}}}}+\arctanh\pa{\frac{\sqrt{\dt r_a^2-\dt r_0^2}}{\dt r_a}}+\frac{1}{2}\log\pa{\frac{1-\tilde{\chi}}{\pa{8\tilde{\chi}}^2}\dt r_0^2}}.
\end{align}
This expression simplifies when the bounds of integration cover the entire range $[\dt r_0,+\infty)$ of allowed radial motion.  Note that the second term vanishes as the lower bound of integration $\dt r_a\to\dt r_0$.  Moreover, the argument of the first term goes to $\sqrt{\tilde{\chi}}$ as $\dt r_b\to\infty$, leaving
\begin{align}
	I_r^\mathrm{nf,out}(\dt r_0,\infty)&=-\frac{1}{2\tilde{r}\sqrt{\tilde{\chi}}}\br{\arctanh{\sqrt{\tilde{\chi}}}+\frac{1}{2}\log\pa{\frac{1-\tilde{\chi}}{\pa{8\tilde{\chi}}^2}\dt r_0^2}}\nonumber\\
	\label{eq:RadialIntegralFullRange}
	&=-\frac{1}{4\tilde{r}\sqrt{\tilde{\chi}}}\log\br{\pa{\frac{1+\sqrt{\tilde{\chi}}}{8\tilde{\chi}}}^2\dt r_0^2}.
\end{align}

For photons arriving inside $\mathcal{C}$, we must separately consider the two regions of the far-zone.  In the right region containing asymptotic infinity, we integrate from a near-zone point $0<\dt r_a\ll 1$ to a far-zone point $\dt r_b\gg\dt r_a>0$, so we label this definite integral ``nf'' for ``near-far''.  The answer is 
\begin{align}
	\label{eq:RadialIntegralNearFar}
	I_r^\mathrm{nf,in}(\dt r_a,\dt r_b)=-\frac{1}{2\tilde{r}\sqrt{\tilde{\chi}}}\br{\arctanh\pa{\frac{\sqrt{\mathcal{Q}(\dt r_b)}}{1+\frac{\dt r_b}{2\tilde{\chi}}}}+\arctanh\pa{\frac{\dt r_a}{\sqrt{\dt r_a^2-\dt r_0^2}}}+\frac{1}{2}\log\pa{\frac{1-\tilde{\chi}}{\pa{8\tilde{\chi}}^2}\ab{\dt r_0^2}}}.
\end{align}
In the left region containing the event horizon, we instead integrate from a far-zone point $\dt r_a<0$ to a near-zone point $\dt r_b<0$, with $\ab{\dt r_b}\ll1$ and $\ab{\dt r_b}\ll\ab{\dt r_a}$, so we label this integration ``fn'' for far-near.  The answer involves a single change of sign,
\begin{align}
	\label{eq:RadialIntegralFarNear}
	I_{r}^\mathrm{fn,in}(\dt r_a,\dt r_b)=-\frac{1}{2\tilde{r}\sqrt{\tilde{\chi}}}\br{\arctanh\pa{\frac{\sqrt{\mathcal{Q}(\dt r_a)}}{1+\frac{\dt r_a}{2\tilde{\chi}}}}-\arctanh\pa{\frac{\dt r_b}{\sqrt{\dt r_b^2-\dt r_0^2}}}+\frac{1}{2}\log\pa{\frac{1-\tilde{\chi}}{\pa{8\tilde{\chi}}^2}\ab{\dt r_0^2}}}.
\end{align}

The calculation proceeds identically for $I_\phi$ and $I_t$, which are conveniently expressed in terms of the $I_r$ integrals:
\begin{align}
	I_\phi^\mathrm{nf,in/out}(\dt r_a,\dt r_b)&=a\pa{\frac{\tilde{r}+M}{\tilde{r}-M}}I_r^\mathrm{nf,in/out}(\dt r_a,\dt r_b)-\frac{aM}{\tilde{r}\sqrt{\tilde{\chi}}}\br{\mathcal{Q}_\phi(\dt r_b)-\mathcal{Q}_\phi(0)},\\
	I_\phi^\mathrm{fn,in}(\dt r_a,\dt r_b)&=a\pa{\frac{\tilde{r}+M}{\tilde{r}-M}}I_r^\mathrm{fn,in}(\dt r_a,\dt r_b)-\frac{aM}{\tilde{r}\sqrt{\tilde{\chi}}}\br{\mathcal{Q}_\phi(\dt r_a)-\mathcal{Q}_\phi(0)},\\
	I_t^\mathrm{nf,in/out}(\dt r_a,\dt r_b)&=\tilde{r}^2\pa{\frac{\tilde{r}+3M}{\tilde{r}-M}}I_r^\mathrm{nf,in/out}(\dt r_a,\dt r_b)-\frac{\tilde{r}}{2\sqrt{\tilde{\chi}}}\br{\mathcal{Q}_t(\dt r_b)-\mathcal{Q}_t(0)},\\
	I_t^\mathrm{fn,in}(\dt r_a,\dt r_b)&=\tilde{r}^2\pa{\frac{\tilde{r}+3M}{\tilde{r}-M}}I_r^\mathrm{fn,in}(\dt r_a,\dt r_b)-\frac{\tilde{r}}{2\sqrt{\tilde{\chi}}}\br{\mathcal{Q}_t(\dt r_a)-\mathcal{Q}_t(0)}.
\end{align}

\subsection{One point in the left far-zone and the other point in the right far-zone}

The last remaining case of relevance is when the geodesic has a lower endpoint $\dt r_a$ in the left far-zone ($\dt r_a<0$) and an upper endpoint $\dt r_b$ in the right far-zone ($\dt r_b>0$).  In this case, the photon passes through the near-zone, and we may obtain the radial integral by adding together the expressions for the near-far and far-near cases derived above.  For $I_r$, summing Eqs.~\eqref{eq:RadialIntegralNearFar}--\eqref{eq:RadialIntegralFarNear} results in
\begin{align}
	I_r^\mathrm{lr}(\dt r_a,\dt r_b)&=I_r^\mathrm{fn,in}(\dt r_a,\dt R)+I_r^\mathrm{nf,in}(\dt R,\dt r_b)\nonumber\\
	\label{eq:RadialIntegralLeftRight}
	&=-\frac{1}{2\tilde{r}\sqrt{\tilde{\chi}}}\br{\arctanh\pa{\frac{\sqrt{\mathcal{Q}(\dt r_a)}}{1+\frac{\dt r_a}{2\tilde{\chi}}}}+\arctanh\pa{\frac{\sqrt{\mathcal{Q}(\dt r_b)}}{1+\frac{\dt r_b}{2\tilde{\chi}}}}+\log\pa{\frac{1-\tilde{\chi}}{\pa{8\tilde{\chi}}^2}\ab{\dt r_0^2}}},
\end{align}
from which the arbitrary radius $\dt R$ has cancelled out.  Here, the label ``lr'' stands for ``left-right''.  Likewise,
\begin{align}
	I_\phi^\mathrm{lr}(\dt r_a,\dt r_b)&=a\pa{\frac{\tilde{r}+M}{\tilde{r}-M}}I_r^\mathrm{lr}(\dt r_a,\dt r_b)-\frac{aM}{\tilde{r}\sqrt{\tilde{\chi}}}\br{\mathcal{Q}_\phi(\dt r_a)+\mathcal{Q}_\phi(\dt r_b)-2\mathcal{Q}_\phi(0)},\\
	I_t^\mathrm{lr}(\dt r_a,\dt r_b)&=\tilde{r}^2\pa{\frac{\tilde{r}+3M}{\tilde{r}-M}}I_r^\mathrm{lr}(\dt r_a,\dt r_b)-\frac{\tilde{r}}{2\sqrt{\tilde{\chi}}}\br{\mathcal{Q}_t(\dt r_a)+\mathcal{Q}_t(\dt r_b)-2\mathcal{Q}_t(0)}.
\end{align}

\subsection{Full answer for \texorpdfstring{$I_r$}{Ir} outside \texorpdfstring{$\mathcal{C}$}{C}}
\label{sec:Outside}

We have now computed all the basic definite integrals that are needed to obtain the full radial integrals $I_r$, $I_\phi$, and $I_t$ for a photon reaching a distant observer at large radius $r_o\to\infty$.  As an example of how to glue them together, we now explicitly consider the radial integral $I_r$.  It is straightforward to similarly assemble $I_\phi$ and $I_t$.

First, consider a photon arriving outside $\mathcal{C}$ (i.e., with $\dt r_0^2>0$).  Tracing back in time from the detector, the photon reaches a radial turning point $\dt r_0$ in the near-zone and then returns to infinity.  Its radial motion in the allowed range $[\dt r_0,+\infty)$ can thus be divided into four stages, as follows.

Before the photon reaches the near-zone, the integral is given by
\begin{align}
	\label{eq:RadialIntegralFarOut}
	I_r\approx I_r^\mathrm{ff}(\dt r_s,\infty)
	=-\frac{1}{2\tilde{r}\sqrt{\tilde{\chi}}}\br{\arctanh\sqrt{\tilde{\chi}}-\arctanh\pa{\frac{\sqrt{\mathcal{Q}(\dt r_s)}}{1+\frac{\dt r_s}{2\tilde{\chi}}}}}.
\end{align}
Once the photon reaches the near-zone, but before it reaches the turning point, the integral is given by the limit $\dt r_o\to\infty$ of Eq.~\eqref{eq:OutsideMatching}:
\begin{align}
	I_r\approx I_r^\mathrm{nf,out}(\dt r_s,\infty)
	=-\frac{1}{2\tilde{r}\sqrt{\tilde{\chi}}}\br{\arctanh\sqrt{\tilde{\chi}}+\arctanh\pa{\frac{\sqrt{\dt r_s^2-\dt r_0^2}}{\dt r_s}}+\frac{1}{2}\log\pa{\frac{1-\tilde{\chi}}{\pa{8\tilde{\chi}}^2}\dt r_0^2}}.
\end{align}
Once the photon reaches the turning point, but before it exits the near-zone, 
\begin{align}
	I_r\approx I_r^\mathrm{nn,out}(\dt r_0,\dt r_s)+I_r^\mathrm{nf,out}(\dt r_0,\infty)
	=-\frac{1}{2\tilde{r}\sqrt{\tilde{\chi}}}\cu{-\arctanh\pa{\frac{\sqrt{\dt r_s^2-\dt r_0^2}}{\dt r_s}}+\frac{1}{2}\log\br{\pa{\frac{1+\sqrt{\tilde{\chi}}}{8\tilde{\chi}}}^2\dt r_0^2}},
\end{align}
where the first term is obtained from Eq.~\eqref{eq:NearIntegralOut} and the second from Eq.~\eqref{eq:RadialIntegralFullRange}.  Once the photon exits the near-zone, the integral is given by
\begin{align}
	I_r&\approx I_r^\mathrm{nf,out}(\dt r_0,\dt r_s)+I_r^\mathrm{nf,out}(\dt r_0,\infty)\nonumber\\
	&=-\frac{1}{2\tilde{r}\sqrt{\tilde{\chi}}}\cu{\arctanh\pa{\frac{\sqrt{\mathcal{Q}(\dt r_s)}}{1+\frac{\dt r_s}{2\tilde{\chi}}}}+\frac{1}{2}\log\pa{\frac{1-\tilde{\chi}}{\pa{8\tilde{\chi}}^2}\dt r_0^2}+\frac{1}{2}\log\br{\pa{\frac{1+\sqrt{\tilde{\chi}}}{8\tilde{\chi}}}^2\dt r_0^2}}.
\end{align}
When the photon finally reaches infinity again, the complete integral is
\begin{align}
	\label{eq:Outside}
	I_r\approx2I_r^\mathrm{nf,out}(\dt r_0,\infty)
	=-\frac{1}{2\tilde{r}\sqrt{\tilde{\chi}}}\log\br{\pa{\frac{1+\sqrt{\tilde{\chi}}}{8\tilde{\chi}}}^2\dt r_0^2}.
\end{align}

\subsection{Full answer for \texorpdfstring{$I_r$}{Ir} inside \texorpdfstring{$\mathcal{C}$}{C}}
\label{sec:Inside}

Now consider a photon arriving inside $\mathcal{C}$ (i.e., with $\dt r_0^2<0$).  Tracing back in time from the detector, the photon passes through the near-zone on its way to the event horizon, never encountering a radial turning point.  Its radial motion in the allowed range $[\dt r_+,+\infty)$ can thus be divided into three stages, as follows.

Before the photon reaches the near-zone, the integral is given by
\begin{align}
	I_r\approx I_r^\mathrm{ff}(\dt r_s,\infty)
	=-\frac{1}{2\tilde{r}\sqrt{\tilde{\chi}}}\br{\arctanh\sqrt{\tilde{\chi}}-\arctanh\pa{\frac{\sqrt{\mathcal{Q}(\dt r_s)}}{1+\frac{\dt r_s}{2\tilde{\chi}}}}},
\end{align}
which is the same expression as outside $\mathcal{C}$, Eq.~\eqref{eq:RadialIntegralFarOut}.  Once the photon enters the near-zone, but before it exits the near-zone, the integral is given by Eq.~\eqref{eq:RadialIntegralNearFar}:
\begin{align}
	I_r\approx I_{r}^\mathrm{nf,in}(\dt r_s,\infty)
	=-\frac{1}{2\tilde{r}\sqrt{\tilde{\chi}}}\br{\arctanh\sqrt{\tilde{\chi}}+\arctanh\pa{\frac{\dt r_s}{\sqrt{\dt r_s^2-\dt r_0^2}}}+\frac{1}{2}\log\pa{\frac{1-\tilde{\chi}}{\pa{8\tilde{\chi}}^2}\ab{\dt r_0^2}}}.
\end{align}
Once the photon exits the near-zone, but before it crosses the horizon, the integral is given by the $\dt r_b\to\infty$ limit of Eq.~\eqref{eq:RadialIntegralLeftRight},
\begin{align}
	I_r\approx I_r^\mathrm{lr}(r_s,\infty)
	=-\frac{1}{2\tilde{r}\sqrt{\tilde{\chi}}}\br{\arctanh\pa{\frac{\sqrt{\mathcal{Q}(\dt r_s)}}{1+\frac{\dt r_s}{2\tilde{\chi}}}}+\arctanh\sqrt{\tilde{\chi}}+\log\pa{\frac{1-\tilde{\chi}}{\pa{8\tilde{\chi}}^2}\ab{\dt r_0^2}}}.
\end{align}
Finally, when the photon crosses the horizon, the complete integral is given by
\begin{align}
	\label{eq:Inside}
	I_r\approx I_r^\mathrm{lr}(r_+,\infty)
	&=-\frac{1}{2\tilde{r}\sqrt{\tilde{\chi}}}\br{\arctanh\pa{\frac{\sqrt{\mathcal{Q}(\dt r_+)}}{1+\frac{\dt r_+}{2\tilde{\chi}}}}+\arctanh{\sqrt{\tilde{\chi}}}+\log\pa{\frac{1-\tilde{\chi}}{\pa{8\tilde{\chi}}^2}\ab{\dt r_0^2}}}\nonumber\\
	&=-\frac{1}{2\tilde{r}\sqrt{\tilde{\chi}}}\log\br{\frac{\sqrt{1-\tilde{\chi}}\pa{1+\sqrt{\tilde{\chi}}}}{\pa{8\tilde{\chi}}^2}\sqrt{\frac{1+\mathcal{Q}_2(\dt r_+,0)}{1-\mathcal{Q}_2(\dt r_+,0)}}\ab{\dt r_0^2}}.
\end{align}
Interestingly, note that the square root containing $\mathcal{Q}_2$ can be pulled out of the logarithm, since
\begin{align}
	\sqrt{\frac{1+\mathcal{Q}_2(\dt r_+,0)}{1-\mathcal{Q}_2(\dt r_+,0)}}=e^{\arctanh{\mathcal{Q}_2(\dt r_+,0)}}.
\end{align}

\subsection{Perpendicular distance from \texorpdfstring{$\mathcal{C}$}{C}}
\label{app:Distance}

The logarithmic approximations for the radial integrals presented thus far are written in terms of the variable $\dt r_0^2$ defined in Eq.~\eqref{eq:Fr} above.  For each choice of $\tilde{r}$, this quantity encodes the arrival positions of photons near the associated point $(\tilde{\alpha},\tilde{\beta})$ on the curve $\mathcal{C}$ [Eq.~\eqref{eq:CurveCoordinates}], expressed in terms of their fractional deviations in conserved quantities $\dt\lambda$ and $\dt q$.  Since the point $\tilde{r}$ is arbitrary, we are in effect using three coordinates $(\tilde{r},\dt\lambda,\dt q)$ to describe positions on a two-dimensional image plane.  A convenient way to remove this large redundancy is to consider only perpendicular displacements from $\mathcal{C}$, denoting the signed distance by $d$ (i.e., $d<0$ inside and $d>0$ outside the closed curve $\mathcal{C}$).  We expect this choice to provide the best approximation for a given point near the curve $\mathcal{C}$, since the line segment intersecting the curve perpendicularly is the shortest.  In this appendix, we relate $\dt r_0^2$ to $d$ [Eq.~\eqref{eq:Distance} below], restricting to perpendicular displacements.  Plugging into the above logarithmic approximations gives the desired expressions in terms of the coordinates $(\tilde{r},d)$ depicted in Fig.~\ref{fig:Screen}.

Since $\dt r_0^2=0$ corresponds to the curve $\mathcal{C}$, the gradient of $\dt r_0^2$ in the image plane $(\alpha,\beta)$ is perpendicular to $\mathcal{C}$.  The norm of the gradient therefore gives the rate of change in the perpendicular direction,
\begin{align}
	\label{eq:IKnewThisWouldComeInHandySomeDay}
	\frac{\dt r_0^2}{d}\approx\ab{\vec{\nabla}\left(\dt r_0^2\right)}_\mathcal{C},
\end{align}
where we restrict $\dt r_0^2$ to perpendicular displacements.  To compute the gradient, we first express $\dt r_0^2$ in terms of $\alpha$ and $\beta$.  Using the inverse of Eq.~\eqref{eq:ScreenCoordinates},
\begin{align}
	\lambda=-\alpha\sin{\theta_o},\qquad
	q=\sqrt{\pa{\alpha^2-a^2}\cos^2{\theta_o}+\beta^2},
\end{align}
one finds that, to leading order in a small deviation from the curve $\mathcal{C}$ with $|\alpha/\tilde{\alpha}-1|\sim|\beta/\tilde{\beta}-1|\ll1$,
\begin{align}
	\dt\lambda=\frac{\lambda}{\tilde{\lambda}}-1
	\approx\frac{\alpha}{\tilde{\alpha}}-1,\qquad
	\dt q=\frac{q}{\tilde{q}}-1
	\approx\frac{\tilde{\alpha}\cos^2{\theta_o}\pa{\alpha-\tilde{\alpha}}+\tilde{\beta}\pa{\beta-\tilde{\beta}}}{\tilde{q}^2}.
\end{align}
Plugging these relations into Eq.~\eqref{eq:Fr} results in
\begin{align}
	\label{eq:psiCritical}
	\dt r_0^2\approx\frac{\Delta(\tilde{r})}{2\tilde{r}^4\tilde{\chi}}\br{\tilde{\psi}\pa{\alpha-\tilde{\alpha}}+\tilde{\beta}\pa{\beta-\tilde{\beta}}},\qquad
	\tilde{\psi}=\tilde{\alpha}-\pa{\frac{\tilde{r}+M}{\tilde{r}-M}}a\sin{\theta_o},
\end{align}
from which we may read off the gradient as
\begin{align}
	\label{eq:Gradient}
	\vec{\nabla}\pa{\dt r_0^2}=\frac{\Delta(\tilde{r})}{2\tilde{r}^4\tilde{\chi}}\pa{\tilde{\psi}\pd_\alpha+\tilde{\beta}\pd_\beta}.
\end{align}
(In light of the flat metric $ds^2=\ed\alpha^2+\ed\beta^2$ on the image plane, the vector fields $\{\pd_\alpha,\pd_\beta\}$ coincide with the unit vectors $\{\hat{\alpha},\hat{\beta}\}$.)  From Eqs.~\eqref{eq:IKnewThisWouldComeInHandySomeDay} and \eqref{eq:Gradient}, it therefore follows that, when $\dt r_0^2$ is evaluated on a perpendicular displacement,
\begin{align}
	\label{eq:Distance}
	d=\frac{2\tilde{r}^4\tilde{\chi}}{\Delta(\tilde{r})}\frac{\dt r_0^2}{\sqrt{\tilde{\beta}^2+\tilde{\psi}^2}}.
\end{align}

Finally, we also present expressions for the unit tangent and normal to $\mathcal{C}$.  The parameter derivatives are given by
\begin{align}
	\label{eq:CurveVelocity}
	\alpha'(\tilde{r})=\frac{2\tilde{r}\tilde{\chi}}{a\sin{\theta_o}}>0,\qquad
	\beta'(\tilde{r})=-\frac{\tilde{\psi}}{\tilde{\beta}}\alpha'(\tilde{r}).
\end{align}
The unit tangent vector to $\mathcal{C}$ is therefore
\begin{align}
	\hat{T}=\pm_o\frac{\alpha'(\tilde{r})\pd_\alpha+\beta'(\tilde{r})\pd_\beta}{\sqrt{\br{\alpha'(\tilde{r})}^2+\br{\beta'(\tilde{r})}^2}}
	=\frac{\tilde{\beta}\pd_\alpha-\tilde{\psi}\pd_\beta}{\sqrt{\tilde{\beta}^2+\tilde{\psi}^2}},
\end{align}
where the inclusion of the sign $\pm_o=\sign\pa{\beta}$ guarantees that $\hat{T}$ points clockwise around $\mathcal{C}$, which corresponds to the direction of increasing/decreasing $\tilde{r}$ in the upper/lower half of the image plane (see Fig.~\ref{fig:Screen}).  As such, the outward normal is obtained by rotating $\hat{T}$ by $90^\circ$ counterclockwise in the image plane:
\begin{align}
	\hat{n}=\frac{\tilde{\psi}\pd_\alpha+\tilde{\beta}\pd_\beta}{\sqrt{\tilde{\beta}^2+\tilde{\psi}^2}}.
\end{align}
We thus confirm directly that the gradient of $\dt r_0^2$ is proportional to $\hat{n}$,
\begin{align}
	\vec{\nabla}\pa{\dt r_0^2}=\frac{\Delta(\tilde{r})}{2\tilde{r}^4\tilde{\chi}}\sqrt{\tilde{\beta}^2+\tilde{\psi}^2}\,\hat{n}.
\end{align}

\subsection{Range of \texorpdfstring{$\mathcal{Q}_2$}{Q2}}
\label{sec:Range}

In this section, we prove that the range of the bivariate function $\mathcal{Q}_2(\dt r_a,\dt r_b)$ defined in Eq.~\eqref{eq:Q2} is $(0,1]$.  This guarantees that the expressions derived in Sec.~\ref{sec:FarFar} for the far-zone integrals are indeed (manifestly) real, as claimed.

We assume that both $\dt r_a$ and $\dt r_b$ are outside the event horizon.  That is, we assume that $\dt r_a\ge\dt r_+$ and $\dt r_b\ge\dt r_+$, though neither $\dt r_a$ nor $\dt r_b$ need be positive.  In that case, we have both $\mathcal{Q}(\dt r_a)>0$ and $\mathcal{Q}(\dt r_b)>0$, since the roots $\dt r_0^\pm=2\pa{-1\pm\sqrt{1-\tilde{\chi}}}$ of $\mathcal{Q}(\dt r)$ always obey
\begin{align}
	\dt r_0^-<\dt r_0^+\le\dt r_-<\dt r_+<0.
\end{align}

First, we wish to prove that $0\le\mathcal{Q}_2(\dt r_a,\dt r_b)$, or equivalently, that
\begin{align}
	0<\mathcal{Q}(\dt r_a)+\mathcal{Q}(\dt r_b)-\pa{\frac{\dt r_a-\dt r_b}{2\sqrt{\tilde{\chi}}}}^2.
\end{align}
Expanding and canceling terms leaves
\begin{align}
	0<2+\frac{1}{\tilde{\chi}}\pa{\dt r_a+\dt r_b+\frac{\dt r_a\dt r_b}{2}}.
\end{align}
In terms of the positive quantities $p_a=\dt r_a-\dt r_0^+>0$ and $p_b=\dt r_b-\dt r_0^+>0$, this reduces to the inequality
\begin{align}
	0<\frac{\sqrt{1-\tilde{\chi}}}{\tilde{\chi}}\pa{p_a+p_b}+\frac{p_ap_b}{2\tilde{\chi}},
\end{align}
which is manifestly satisfied since $p_a$, $p_b$, $\tilde{\chi}$, and $\sqrt{1-\tilde{\chi}}$ are all positive.

Next, we need to show that $\mathcal{Q}_2(\dt r_a,\dt r_b)\le1$, or equivalently, that
\begin{align}
	2\sqrt{\mathcal{Q}(\dt r_a)}\sqrt{\mathcal{Q}(\dt r_b)}\le\mathcal{Q}(\dt r_a)+\mathcal{Q}(\dt r_b)-\pa{\frac{\dt r_a-\dt r_b}{2\sqrt{\tilde{\chi}}}}^2.
\end{align}
Completing the square and rearranging yields
\begin{align}
	\pa{\frac{\dt r_b-\dt r_a}{2\sqrt{\tilde{\chi}}}}^2\le\br{\sqrt{\mathcal{Q}(\dt r_b)}-\sqrt{\mathcal{Q}(\dt r_a)}}^2.
\end{align}
From now on, we assume without loss of generality that $\dt r_b\ge\dt r_a$.  Then taking the square root of both sides leaves
\begin{align}
	0\le\frac{\dt r_b-\dt r_a}{2\sqrt{\tilde{\chi}}}
	\le\sqrt{\mathcal{Q}(\dt r_b)}-\sqrt{\mathcal{Q}(\dt r_a)},
\end{align}
or equivalently,
\begin{align}
	0\le\pa{\sqrt{\mathcal{Q}(\dt r_b)}-\frac{\dt r_b}{2\sqrt{\tilde{\chi}}}}-\pa{\sqrt{\mathcal{Q}(\dt r_a)}-\frac{\dt r_a}{2\sqrt{\tilde{\chi}}}}.
\end{align}
A simple way to establish that this inequality holds is by noting that the function
\begin{align}
	\mathcal{P}(\dt r)=\sqrt{\mathcal{Q}(\dt r)}-\frac{\dt r}{2\sqrt{\tilde{\chi}}}
\end{align}
is monotonic on the radial range $\dt r>\dt r_+$ of interest.  Indeed, letting $p=\dt r-\dt r_0^+>0$, one finds that
\begin{align}
	\mathcal{P}'(\dt r)=\frac{2\sqrt{1-\tilde{\chi}}-\sqrt{\tilde{\chi}}+p}{2\sqrt{\tilde{\chi}}}
	>\frac{2\sqrt{1-\tilde{\chi}}-\sqrt{\tilde{\chi}}}{2\sqrt{\tilde{\chi}}}
	>0,
\end{align}
where the last inequality follows from the range of $\tilde{\chi}\in(0,3/4]$.

Finally, note that $\mathcal{Q}_2(\dt r,\dt r)=1$ for all $\dt r$, so the upper bound may be saturated.  On the other hand, the lower bound may not, since $\mathcal{Q}_2(\dt r,x)=0$ if and only if $x=\dt r_0^\pm<\dt r_+$, which is outside the range of $\dt r$ under consideration.

\bibliography{KerrRing}
\bibliographystyle{utphys}

\end{document}